\documentclass[utf8]{frontiersFPHY} 

\usepackage{url,hyperref,microtype,subcaption}
\usepackage[onehalfspacing]{setspace}
\usepackage{amsmath}
\usepackage{bbold}
\usepackage{physics}

\def\keyFont{\fontsize{8}{11}\helveticabold }
\def\firstAuthorLast{Holt {et~al.}} 
\def\Authors{Jeremy W.\ Holt\,$^{1,2,*}$, Mamiya Kawaguchi\,$^{3,4}$ and Norbert Kaiser\,$^{5}$}
\def\Address{$^{1}$Department of Physics and Astronomy, Texas A\&M University, College Station, TX, USA \\
$^{2}$ Cyclotron Institute, Texas A\&M University, College Station, TX, USA \\
$^{3}$Department of Physics, Nagoya University, Nagoya, Japan \\
$^{4}$Department of Physics, Center for Particle Physics and Field Theory, Fudan University, Shanghai, China \\
$^{5}$Physik-Department, Technische Universit\"at M\"unchen, Garching, Germany
}
\def\corrAuthor{Jeremy W.\ Holt}

\def\corrEmail{holt@physics.tamu.edu}

\begin{document}
\onecolumn
\firstpage{1}

\title[Implementing chiral three-body forces]{Implementing chiral three-body forces in terms of medium-dependent two-body forces} 

\author[\firstAuthorLast ]{\Authors} 
\address{} 
\correspondance{} 

\extraAuth{}

\maketitle

\begin{abstract}
Three-nucleon (3N) forces are an indispensable ingredient for accurate 
few-body and many-body nuclear structure and reaction theory calculations. 
While the direct implementation of chiral 3N forces can be technically very challenging, a 
simpler approach is given by employing instead a medium-dependent 
NN interaction $V_\text{med}$ that reflects the physics of three-body forces at
the two-body normal-ordered approximation. 
We review the derivation and construction of $V_\text{med}$ from the 
chiral 3N interaction at next-to-next-to-leading order (N2LO), 
consisting of a long-range $2\pi$-exchange term, a mid-range $1\pi$-exchange component 
and a short-range contact-term. Several applications of $V_\text{med}$ 
to the equation of state of cold nuclear and neutron matter, the nucleon
single-particle potential in nuclear matter, 
and the nuclear quasiparticle interaction are discussed. We also 
explore differences in using local vs.\ nonlocal regulating functions on 3N 
forces and make direct comparisons to exact results at low order in perturbation theory 
expansions for the equation of state and
single-particle potential. We end with a discussion and numerical calculation of the 
in-medium NN potential $V_\text{med}$ from the next-to-next-to-next-to-leading
order (N3LO) chiral 3N force, 
which consists of a series of long-range and short-range terms.



\tiny
 \keyFont{ \section{Keywords:} Chiral effective field theory, three-body forces, nuclear matter, equation of state, nuclear reaction theory} 
\end{abstract}

\section{Introduction}

Three-nucleon forces are essential to any microscopic description of nuclear many-body systems, from the 
structure and reactions of finite nuclei \cite{Hergert2013,Hebeler:2015,carlson15,Navratil:2016} 
to the equation of state and transport properties of dense matter 
encountered in core-collapse supernovae and neutron stars 
\cite{Hebeler:2015,wiringa88,baldo97,bogner05,coraggio13,gezerlis13,Baardsen2013,carbone14,sammarruca15,drischler16}. 
Three-body forces have been shown to dramatically improve the saturation properties of nuclear matter
\cite{baldo97,bogner05,Akmal98}, though there are still large uncertainties compared to the empirical 
saturation energy and density. Three-nucleon forces also now being routinely implemented in a number of 
ab initio many-body methods such as the no-core shell model \cite{Barrett2013}, coupled-cluster theory
\cite{Hagen2012A,Hagen:2013nca}, self-consistent Green's function theory \cite{Soma2013}, the
similarity renormalization group \cite{Roth2011,Stroberg:2019mxo}, and quantum Monte Carlo \cite{carlson15} 
to study nuclear ground-state and excited states up to medium-mass nuclei. In particular, three-body forces have been 
shown to be especially relevant for understanding the properties of neutron-rich nuclei out to the drip line
\cite{Hagen2012A,Soma2013,Otsuka2010,Hagen2012B}. 

In the past, it has been challenging \cite{wiringa88} to obtain nuclear two- and three-body forces that simultaneously
fit well the properties of finite nuclei and nuclear matter, but 
in recent years, much progress has been achieved within the framework of chiral effective field theory
\cite{WEINBERG79,epelbaum09rmp,MACHLEIDT11,entem17,reinert18} to 
construct three-body forces consistent with the employed two-body force, all within a systematic power 
series expansion involving the ratio of the physical scale $Q$ to the chiral symmetry breaking scale 
$\Lambda_\chi \sim 1$\,GeV.
In chiral effective field theory with explicit nucleon and pion degrees of freedom only, three-nucleon forces appear
first at third order in the chiral expansion $(Q/\Lambda_\chi)^3$, or next-to-next-to-leading (N2LO) order. These 
leading contributions to the chiral
three-nucleon force (3NF) are now routinely employed in nuclear structure and reaction theory calculations, 
but many-body 
contributions at N3LO \cite{ishikawa07,bernard08,bernard11,epelbaum06} are expected to be important.

In the present work, we will review how to implement three-nucleon forces via medium-dependent 
two-body interactions \cite{baldo97,bogner05,hagen07} 
in nuclear many-body calculations of the equation of state, single-particle potential,
and quasiparticle interaction. We will show that this approach provides an excellent approximation at first
order in many-body perturbation theory by comparing to exact results from three-body forces. At higher orders
in perturbation theory, the use of medium-dependent NN interactions fail to reproduce all topologies, however,
residual three-body interactions have been shown \cite{drischler19} to give relatively small contributions ($\sim 1$\,MeV)
to the nuclear equation of state at saturation density up to second order in perturbation theory. We also consider
several commonly used high-momentum regulating functions for three-body forces and study their impact on 
the density-dependent interaction $V_{\rm med}$. In particular, we find that local regulators introduce large
artifacts compared to nonlocal regulators when the same value of the cutoff scale $\Lambda$ is used 
in both cases.

\section{From three-body forces to medium-dependent two-body forces}
\subsection{Chiral three-body force at next-to-next-to-leading order}

The nuclear Hamiltonian can generically be written in the form
\begin{equation}
H = \sum_i \frac{\vec p_i^{\,2}}{2M} + \frac{1}{2} \sum_{ij} V_{ij} + \frac{1}{6} \sum_{ijk} V_{ijk} + \cdots,
\end{equation}
where $\vec p_i$ is the momentum of nucleon $i$, $V_{ij}$ represents the two-body interaction between particles
$i$ and $j$, and $V_{ijk}$ represents the three-body interaction between particles $i$, $j$, $k$.
\begin{figure}[t]
\begin{center}
\includegraphics[height=3.cm]{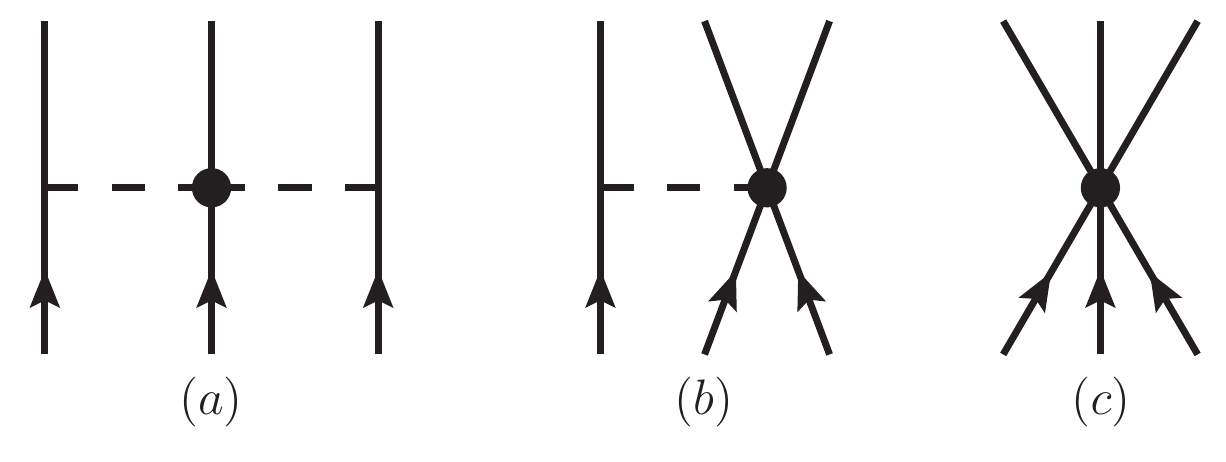}
\end{center}
\caption{Diagrammatic contributions to the chiral three-nucleon force at next-to-next-to-leading order (N2LO)
in the chiral expansion.}
\label{fig:3NF}
\end{figure}
Three-body forces emerge first at N2LO in the chiral expansion and contain three different topologies
represented diagrammatically in Fig.\ \ref{fig:3NF}. The 
two-pion-exchange three-body force, Fig.\ \ref{fig:3NF}(a), contains terms proportional to the
low-energy constants $c_1$, $c_3$, and $c_4$:
\begin{equation}
V_{3N}^{(2\pi)} = \sum_{i\neq j\neq k} \frac{g_A^2}{8f_\pi^4}
\frac{\vec{\sigma}_i \cdot \vec{q}_i \, \vec{\sigma}_j \cdot
\vec{q}_j}{(\vec{q_i}^2 + m_\pi^2)(\vec{q_j}^2+m_\pi^2)}
F_{ijk}^{\alpha \beta}\tau_i^\alpha \tau_j^\beta,
\label{3n1}
\end{equation}
where $g_A = 1.29$ is the axial coupling constant, $m_\pi = 138$\,MeV is the average pion mass, 
$f_\pi = 92.2$\,MeV is the pion decay constant, $\vec{q}_i=\vec k^\prime_i - \vec k_i$ is the change in momentum of
nucleon $i$ (i.e., the momentum transfer), and the isospin tensor $F_{ijk}^{\alpha \beta}$ is defined by
\begin{equation}
F_{ijk}^{\alpha \beta} = \delta^{\alpha \beta}\left (-4c_1m_\pi^2
 + 2c_3 \,\vec{q}_i \cdot \vec{q}_j \right ) +
c_4 \,\epsilon^{\alpha \beta \gamma} \tau_k^\gamma \,\vec{\sigma}_k
\cdot \left ( \vec{q}_i \times \vec{q}_j \right ).
\label{3n4}
\end{equation}
The three low-energy constants $c_1$, $c_3$, and $c_4$ can be fitted to empirical pion-nucleon
\cite{buttiker00,hoferichter15} or nucleon-nucleon \cite{rentmeester03,entem03} scattering data.

The one-pion exchange term, Fig.\ \ref{fig:3NF}(b), proportional to the low-energy constant $c_D$ has the form
\begin{equation}
V_{3N}^{(1\pi)} = -\sum_{i\neq j\neq k} \frac{g_A c_D}{8f_\pi^4\, \Lambda_\chi}
\,\frac{\vec{\sigma}_j \cdot \vec{q}_j}{\vec{q_j}^2+m_\pi^2}\, \vec{\sigma}_i \cdot
\vec{q}_j \, {\vec \tau}_i \cdot {\vec \tau}_j \, ,
\label{3n2}
\end{equation}
where the high-momentum scale is typically taken as $\Lambda_\chi = 700$\,MeV.
The three-nucleon contact force, Fig.\ \ref{fig:3NF}(c), proportional to $c_E$ is written:
\begin{equation}
V_{3N}^{(\rm ct)} = \sum_{i\neq j\neq k} \frac{c_E}{2f_\pi^4\, \Lambda_\chi}
\,{\vec \tau}_i \cdot {\vec \tau}_j\,.
\label{3n3}
\end{equation}
There are several different experimental observables commonly used for fitting the low-energy constants 
$c_D$ and $c_E$. Most approaches fit the binding energies of $A=3$ nuclei together with one of the 
following observables: (a) the neutron-deuteron doublet scattering length \cite{epelbaum02,piarulli18}, 
(b) the radius of $^4$He \cite{hebeler11}, (d) the properties of light and medium-mass nuclei 
\cite{navratil07a,ekstrom15}, and (d) the triton lifetime \cite{gardestig06,gazit09}.

Since the three-nucleon force $V_{3N}$ is symmetric under the interchange of particle labels, there
are only three independent terms from the $i, j, k$ permutations, which allows us to write $V_{3N} = W_1+ W_2+ W_3$. 
For instance, $W_1^{(\rm ct)} = \frac{c_E}{f_\pi^4\, \Lambda} \,{\vec \tau}_2 \cdot {\vec \tau}_3$.
The antisymmetrized three-body
interaction $\bar V_{3N}$ can be written in terms of two-body antisymmetrization operators $P_{ij}$ as follows:
\begin{equation}
\bar V_{3N} = (1-P_{12})(1-P_{13}-P_{23})V_{3N},
\end{equation}
where 
\begin{equation}
P_{ij} = \left ( \frac{1+\vec \sigma_i \cdot \vec \sigma_j}{2} \right ) \left ( \frac{1+\vec \tau_i \cdot \vec \tau_j}{2} \right ),
\hspace{.1in} \vec k_i \longleftrightarrow \vec k_j.
\end{equation}

\subsection{Density-dependent NN interaction at order N2LO}

In second quantization, a three-body force $V_{3N}$ can be written as
\begin{equation}
V_{3N} = \frac{1}{36} \sum_{123456} \langle 1 2 3 | \bar V_{3N} | 4 5 6 \rangle
\hat a^\dagger_1 \hat a^\dagger_2 \hat a^\dagger_3 \hat a_6 \hat a_5 \hat a_4
\label{v3nsq}
\end{equation}
where $\bar V_{3N}$ denotes the antisymmetrized three-body matrix element, and 
$a_i^\dagger$ ($a_i$) are the usual creation (annihilation) operators associated
with state $| i \rangle$.
A medium-dependent two-body interaction can then be constructed by normal ordering the three-body
force with respect to a convenient reference state, such as the ground state of the noninteracting
many-body system, rather than the true vacuum as in Eq.\ (\ref{v3nsq}). Normal ordering with respect
to the noninteracting ground state then produces a three-body force of the form
\begin{eqnarray}
V_{3N} &=& \frac{1}{6} \sum_{ijk}\langle ijk | \bar V_{3N} | ijk \rangle 
+ \frac{1}{2} \sum_{ij1 4} \langle ij1 | \bar V_{3N} | ij4 \rangle :\!\hat a^\dagger_1 \hat a_4\!:
+ \frac{1}{4}\sum_{i1245}\langle i12 | \bar V_{3N} | i 4 5 \rangle :\!\hat a^\dagger_1 \hat a^\dagger_2 
\hat a_5 \hat a_4\!:  \nonumber \\
&& + {1\over 36} \sum_{123456} \langle 1 2 3 | \bar V_{3N} | 4 5 6 \rangle
:\!\hat a^\dagger_1 \hat a^\dagger_2 \hat a^\dagger_3 \hat a_6 \hat a_5 \hat a_4\!:,
\label{nord}
\end{eqnarray}
where $:\! \mathcal{\hat O}\!:$ denotes normal ordering of operator $\mathcal{\hat O}$.
In practice the construction of the medium-dependent two-body force 
\begin{equation}
\frac{1}{4}\sum_{i1245}\langle i12 | \bar V_{3N} | i 4 5 \rangle :\!\hat a^\dagger_1 \hat a^\dagger_2 
\hat a_5 \hat a_4\!:
\end{equation}
then amounts to summing the third particle over the filled 
states in the noninteracting Fermi sea, involving spin and isospin summations as well as momentum integration:
\begin{equation}
\bar V_{\rm med} = \sum_{s_3 t_3} \int \frac{d^3k_3}{(2\pi)^3} \theta(k_f-k_3) (1-P_{13}-P_{23}) V_{3N},
\label{ddnn}
\end{equation}
where $k_f$ is the Fermi momentum and 
we have absorbed the particle exchange operator $(1-P_{12})$ into the definition of the antisymmetrized 
medium-dependent NN interaction $\bar V_{\rm med}$. In general, there are nine different diagrams 
that need to be evaluated independently: $(1-P_{13}-P_{23})(W_1+W_2+W_3)$, which correspond to different closings 
of one incoming and outgoing particle line.

\begin{figure}[t]
\begin{center}
\includegraphics[height=2.6cm]{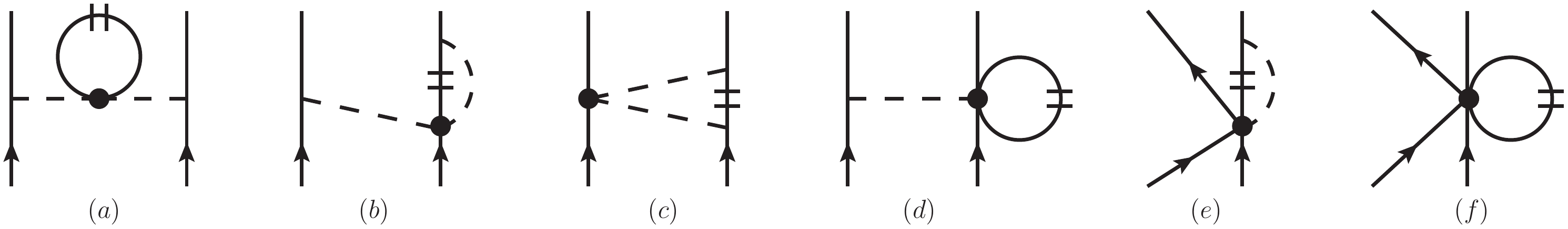}
\end{center}
\caption{Diagrammatic contributions to the density-dependent NN interaction derived from the N2LO
chiral three-nucleon force.}
\label{fig:mednn}
\end{figure}

As a simple example, we compute the density-dependent NN interaction arising from the three-body contact term
at N2LO shown diagrammatically in Fig.\ \ref{fig:mednn} ($f$).
We begin by evaluating the spin and isospin traces in Eq.\ (\ref{ddnn}):
\begin{eqnarray}
&& \Tr_{\sigma_3\tau_3}\large [(1-P_{13}-P_{23})(\vec \tau_2 \cdot \vec \tau_3 + \vec \tau_1 \cdot \vec \tau_3
+ \vec \tau_1 \cdot \vec \tau_2) \large ]\\ \nonumber
&=& \Tr_{\sigma_3\tau_3} (\vec \tau_2 \cdot \vec \tau_3 + \vec \tau_1 \cdot \vec \tau_3
+ \vec \tau_1 \cdot \vec \tau_2) - \Tr_{\sigma_3\tau_3} \left [ \left ( \frac{1+\vec \sigma_1 \cdot \vec \sigma_3}{2}
\frac{1+\vec \tau_1 \cdot \vec \tau_3}{2}  \right )(\vec \tau_2 \cdot \vec \tau_3 + \vec \tau_1 \cdot \vec \tau_3
+ \vec \tau_1 \cdot \vec \tau_2) \right ]\\ \nonumber
&-& \Tr_{\sigma_3\tau_3} \left [ \left ( \frac{1+\vec \sigma_2 \cdot \vec \sigma_3}{2}
\frac{1+\vec \tau_2 \cdot \vec \tau_3}{2}  \right )(\vec \tau_2 \cdot \vec \tau_3 + \vec \tau_1 \cdot \vec \tau_3
+ \vec \tau_1 \cdot \vec \tau_2) \right ]\\ \nonumber
&=& 4 \vec \tau_1 \cdot \vec \tau_2 - \frac{1}{4} \Tr_{\sigma_3\tau_3} ( \vec \tau_1 \cdot \vec \tau_2 + 
(\vec \tau_1 \cdot \vec \tau_3)(\vec \tau_2 \cdot \vec \tau_3) + 
(\vec \tau_1 \cdot \vec \tau_3)(\vec \tau_1 \cdot \vec \tau_3)) \\ \nonumber 
&-& \frac{1}{4} \Tr_{\sigma_3\tau_3} ( \vec \tau_1 \cdot \vec \tau_2
+ (\vec \tau_2 \cdot \vec \tau_3)(\vec \tau_2 \cdot \vec \tau_3) + 
(\vec \tau_2 \cdot \vec \tau_3)(\vec \tau_1 \cdot \vec \tau_3)) \\ \nonumber
&=& 2 \vec \tau_1 \cdot \vec \tau_2 - \frac{1}{2} (4 \vec \tau_1 \cdot \vec \tau_2 + 12 ) = -6,
\end{eqnarray}
where we have used the well known properties of Pauli matrices: 
$\Tr \vec \sigma = 0$, $\Tr \mathbb{1} = 2$, $\Tr_3 (\vec \tau_i \cdot \vec \tau_3)(\vec \tau_j \cdot \vec \tau_3)
= 2 \vec \tau_i \cdot \vec \tau_j$, and $\vec \tau_i \cdot \vec \tau_i = 3$. The integration over filled momentum
states is trivial:
\begin{equation}
\int \frac{d^3k_3}{(2\pi)^3} \theta(k_f-k_3) \frac{c_E}{f_\pi^4\Lambda_\chi} = \frac{1}{2\pi^2}\frac{k_f^3}{3}
\frac{c_E}{f_\pi^4\Lambda_\chi},
\end{equation}
which gives a final result of
\begin{equation}
V_{\rm med}^{(\rm ct)} = -\frac{c_E k_f^3}{\pi^2 f_\pi^4 \Lambda_\chi}.
\label{med6}
\end{equation}
This particularly simple three-body contact interaction gives rise to a momentum-independent effective
two-body interaction. For the more complicated 1$\pi$- and 2$\pi$-exchange topologies, it is convenient
to consider the on-shell scattering ($|\vec p\, | = |\vec p^{\,\prime}|$) of two nucleons in the center-of-mass 
frame: $N_1(\vec p\,) + N_2(-\vec p\,) \rightarrow N_1(\vec p^{\,\prime}) + N_2(-\vec p^{\,\prime})$.
This assumption results in a medium-dependent 2N 
interaction with the same isospin and spin structures as the free-space 2N potential, which allows for a
simple decomposition of $V_{\rm med}$ into partial-wave matrix elements as we show in Section \ref{pwd}. 
In the more general case $N_1(\vec p_1) + N_2(\vec p_2) \rightarrow N_1(\vec p_3) + N_2(\vec p_4)$, the 
in-medium 2N interaction will contain operator structures depending on the center-of-mass
momentum $\vec P = \vec p_1 + \vec p_2 = \vec p_3 + \vec p_4$. Such contributions have been shown to be
small in practice \cite{hebeler10}. In the applications discussed below, higher-order
perturbative contributions to the ground state energy and single-particle energies involve also off-shell matrix 
elements of the interaction $\langle \vec p^{\, \prime} | V | \vec p \, \rangle$, where $|\vec p\, | \ne |\vec p^{\,\prime}|$.
In such cases we use the approximation $p^2 \rightarrow \frac{1}{2}(p^2 + {p^{\prime \,2}})$
in the formulas derived below. The resulting interaction can then be straightfowardly implemented into modern 
nuclear structure codes. We will explicitly test some of the approximations noted above by comparing exact results
at low order in perturbation theory using the full 3-body force to the results using instead the medium-dependent 
2N interaction. 

Note that in the above derivation of the density-dependent 2N interaction associated with $V_{3N}^{(\rm ct)}$, 
we have not applied a high-momentum regulator, which would be necessary to eliminate the components of the nuclear 
interaction that lie beyond the breakdown scale of the effective field theory. In the case of nucleon-nucleon potentials, 
the cutoff scale
is typically chosen $\Lambda \lesssim 700$\,MeV, beyond which the introduction of a new dynamical degree of freedom 
(the $\rho$ meson with mass $m_\rho = 770$\,MeV) would be required. 
On the other hand, in order to fit empirical nucleon-nucleon scattering phase shift data
up to laboratory energies of $E_{\rm lab} = 350$\,MeV, the cutoff is normally chosen $\Lambda \gtrsim 414$\,MeV 
(the relative momentum in the center-of-mass frame corresponding to $E_{\rm lab} = 350$\,MeV). In practice, it is
found that relatively low values of the momentum-space cutoff $\Lambda \lesssim 500$\,MeV lead to perturbative 
nucleon-nucleon potentials, which are suitable for a wide range of methods to solve the quantum 
many-body problem. Such low-cutoff potentials, however, exhibit larger artifacts in calculations of the 
density-dependent ground state energy of nuclear matter and single-particle potential as we will discuss
explicitly below. While the choice of cutoff scale is well motivated, the regulating function
can take various forms. Traditionally, an exponential regulator in the incoming and outgoing relative momenta
is chosen:
\begin{equation}
f(p',p) = \exp[-(p'/\Lambda)^{2n} - (p/\Lambda)^{2n}],
\label{nlr}
\end{equation}
where $\vec p = \frac{1}{2}( \vec p_1 - \vec p_2)$ and $\vec p^{\, \prime} = \frac{1}{2}(\vec p_3 
- \vec p_4 )$ for the general 
two-body scattering process $N(\vec p_1) + N(\vec p_2) \rightarrow N(\vec p_3) + N(\vec p_4)$, and 
$n$ is an integer chosen such that the regulator affects only high powers in the chiral expansion.
More recently \cite{reinert18,epelbaum15}, the pion-exchange components of the nucleon-nucleon
interaction have been regularized in coordinate space according to
\begin{equation}
V_\pi(\vec r \,) \left [ 1 - e^{-r^2/R^2} \right ],
\end{equation}
where $0.8\,{\rm fm} \leq R \leq 1.2\,{\rm fm}$, while the contact terms in the nuclear potential were
regularized according to Eq.\ (\ref{nlr}) above. In previous calculations \cite{holt10,wellenhofer14} 
of the medium-dependent 2N force $V_{\rm med}$, we have imposed the nonlocal regulating function 
above only after the momentum-space integration over $k_3$ is performed. 
This choice led to simplified analytical expressions for the density-dependent NN interaction 
in cold nuclear matter. A three-body regulator that treats all particles symmetrically can be defined 
by \cite{navratil07}
\begin{equation}
W_3 \longrightarrow W_3 F(q_1,q_2) = W_3 \exp[-(q_1/\Lambda)^{4} - (q_2/\Lambda)^{4}],
\label{lr}
\end{equation}
where $\vec q_1 = \vec p_1^{\, \prime} - \vec p_1$ and $\vec q_2 = \vec p_2^{\, \prime} - \vec p_2$ 
are the momentum transfers for particles $1$ and $2$ in $W_3$. Analogous expressions hold for the 
contributions $W_1 \longrightarrow W_1 F(q_2,q_3)$ and $W_2 \longrightarrow W_2 F(q_1,q_3)$. This 
choice of regulating function leads to more complicated expressions for the 
density-dependent 2N interaction since now the regulator in general can involve the momentum $k_3$ over
which we integrate. More importantly, the local regulator in Eq.\ (\ref{lr}) 
leads to much stronger cutoff artifacts for the same choice of scale ($\Lambda_{\rm loc} = \Lambda_{\rm nonloc}$) 
as we will demonstrate in the following. Additional discussion regarding the role of cutoff artifacts
on nuclear many-body calculations can be found in Ref.\ \cite{dyhdalo16}.

To start, when we employ the local regulator in Eq.\ (\ref{lr}), we now find for the density-dependent NN interaction
in isospin-symmetric nuclear matter:
\begin{equation}
V_{\rm med}^{(\rm ct)} = \frac{c_E}{2\pi^2 f_\pi^4 \Lambda_\chi} \left [\frac{4}{3}k_f^3\, \vec \tau_1 \cdot \vec \tau_2
F^2(q^2,\Lambda) -2 \vec \tau_1 \cdot \vec \tau_2 F(q^2,\Lambda) \tilde \Gamma_4(p) - 3 \Gamma^\prime_4(p,q)
\right ],
\end{equation}
where 
\begin{equation}
F(q^2,\Lambda) = e^{-q^4/\Lambda^4},
\end{equation}
\begin{equation}
\tilde \Gamma_4(p) = \int_0^{k_f} dk \int_{-1}^1 dx\, k^2 \, F(p^2 + k^2 + 2pkx,\Lambda),
\label{gam4t}
\end{equation}
\begin{eqnarray}
\Gamma_4^\prime(p,q) &=& \int_0^{k_f} dk \int_{-1}^1 dx \int_{0}^{2\pi} d\phi\, \frac{k^2}{2\pi}
F(p^2 + k^2 + k x \sqrt{4p^2-q^2} + q k \sqrt{1-x^2} \cos \phi,\Lambda) \nonumber \\ 
&&\times F(p^2 + k^2 + k x \sqrt{4p^2-q^2} - q k \sqrt{1-x^2} \cos \phi,\Lambda).
\label{gam4p}
\end{eqnarray}
In the limit of large $\Lambda$ we find that $F(q^2,\Lambda) \rightarrow 1$, 
$\tilde \Gamma_4(p) \rightarrow \frac{2k_f^3}{3}$, and
$\Gamma_4^\prime(p,q) \rightarrow \frac{2k_f^3}{3}$. Thus, in this limit we clearly recover Eq.\ (\ref{med6}).

Previously, for the in-medium pion self-energy correction, Fig.\ \ref{fig:mednn}(a), with no regulator we found
\begin{equation}
V_{NN}^{\rm med, 1} = \frac{g_A^2 k_f^3}{3\pi^2 f_\pi^4} \vec \tau_1 \cdot \vec \tau_2 \frac{\vec \sigma_1 \cdot \vec q
\, \vec \sigma_2 \cdot \vec q}{(m_\pi^2+q^2)^2}\left ( 2c_1m_\pi^2+c_3q^2  \right ).
\end{equation}
With the local regulator in Eq.\ (\ref{lr}) we now find
\begin{equation}
V_{NN}^{\rm med, 1} = \frac{g_A^2 k_f^3}{3\pi^2 f_\pi^4} \vec \tau_1 \cdot \vec \tau_2 \frac{\vec \sigma_1 \cdot \vec q
\, \vec \sigma_2 \cdot \vec q}{(m_\pi^2+q^2)^2}\left ( 2c_1m_\pi^2+c_3q^2  \right ) F^2(q^2,\Lambda).
\end{equation}
Previously, for the Pauli-blocked vertex correction, Fig.\ \ref{fig:mednn}(b), we found
\begin{eqnarray}
&& V_{NN}^{\rm med, 2} = \frac{g_A^2}{8\pi^2 f_\pi^4} \vec \tau_1 \cdot \vec \tau_2 \frac{\vec \sigma_1 \cdot \vec q
\, \vec \sigma_2 \cdot \vec q}{m_\pi^2+q^2}\Big ( -4 c_1 m_\pi^2 [ \Gamma_0 + \Gamma_1 ] -(c_3+c_4)[ q^2 ( \Gamma_0 + 2\Gamma_1 + \Gamma_3 )+4\Gamma_2] \nonumber \\
 &&\hspace{.5in} +  4c_4 \Big [ \frac{2}{3}k_f^3 - m_\pi^2\Gamma_0 \Big ] \Big).
\label{pbvc}
\end{eqnarray}
When the local regulators are employed, we now find that the $p$-dependent auxiliary functions $\Gamma_i$ must be replaced
by
\begin{equation}
\tilde \Gamma_0(p) = \int_0^{k_f} dk \int_{-1}^1 dx\, \frac{k^2}{m_\pi^2 + p^2 + k^2 + 2pkx}F(p^2 + k^2 + 2pkx,\Lambda),
\end{equation}
\begin{equation}
\tilde \Gamma_1(p) = \int_0^{k_f} dk \int_{-1}^1 dx\, \frac{k^3\, x / p}{m_\pi^2 + p^2 + k^2 + 2pkx}F(p^2 + k^2 + 2pkx,\Lambda),
\end{equation}
\begin{equation}
\tilde \Gamma_2(p) = \int_0^{k_f} dk \int_{-1}^1 dx\, \frac{k^4(1-x^2)/2}{m_\pi^2 + p^2 + k^2 + 2pkx}F(p^2 + k^2 + 2pkx,\Lambda),
\end{equation}
\begin{equation}
\tilde \Gamma_3(p) = \int_0^{k_f} dk \int_{-1}^1 dx\, \frac{k^4(3x^2-1)/(2p^2)}{m_\pi^2 + p^2 + k^2 + 2pkx}F(p^2 + k^2 + 2pkx,\Lambda).
\end{equation}
where the unprimed versions of these functions in Eq.\ (\ref{pbvc}) can be obtained by setting $\Lambda \rightarrow \infty$.
In addition, the term $4 c_4 \left[ \frac{2k_f^3}{3} \right ]$ in $V_{NN}^{\rm med,2}$ must be replaced with
the quantity
\begin{equation}
4 c_4 \left[ \frac{2k_f^3}{3} \right ] \longrightarrow 4 c_4 \int_0^{k_f} dk \int_{-1}^1 dx\, k^2 \,
F(p^2 + k^2 + 2pkx,\Lambda) \equiv 4c_4 \tilde \Gamma_4(p).
\end{equation}
Then the revised Pauli-blocked vertex correction has the form
\begin{eqnarray}
&& V_{NN}^{\rm med, 2} = \frac{g_A^2}{8\pi^2 f_\pi^4} \vec \tau_1 \cdot \vec \tau_2 \frac{\vec \sigma_1 \cdot \vec q
\, \vec \sigma_2 \cdot \vec q}{m_\pi^2+q^2}\left ( -4 c_1 m_\pi^2 [ \tilde \Gamma_0 + \tilde \Gamma_1 ] -(c_3+c_4)[ q^2 ( \tilde \Gamma_0 + 2\tilde \Gamma_1 + \tilde \Gamma_3 )+4\tilde \Gamma_2] \right . \nonumber \\
&&\hspace{.5in} \left. +4c_4\left[\tilde \Gamma_4 - m_\pi^2\tilde \Gamma_0 \right ]\right ) F(q^2,\Lambda).
\end{eqnarray}
Previously, we found for the Pauli-blocked two-pion-exchange interaction, Fig.\ \ref{fig:mednn}(c),
\begin{eqnarray}
&&\hspace{-.2in} V_{NN}^{\rm med, 3} = \frac{g_A^2}{16\pi^2 f_\pi^4} \{ -12c_1m_\pi^2\left[ 2\Gamma_0 
- (2m_\pi^2+q^2) G_0 \right ] - c_3[8k_f^3-12(2m_\pi^2+q^2)\Gamma_0 -6q^2\Gamma_1 \nonumber \\
&&+ 3(2m_\pi^2+q^2)^2G_0 ] +4c_4 \vec \tau_1 \cdot \vec \tau_2 ( \vec \sigma_1 \cdot \vec \sigma_2 q^2
- \vec \sigma_1 \cdot \vec q \, \vec \sigma_2 \cdot \vec q )G_2-(3c_3+c_4 \vec \tau_1 \cdot \vec \tau_2 ) 
i (\vec \sigma_1 + \vec \sigma_2) \cdot (\vec q \times \vec p) \nonumber \\ 
&&\times [2\Gamma_0+2\Gamma_1 -(2m_\pi^2 + q^2)(G_0+2G_1) ]
 - 12 c_1 m_\pi^2 i (\vec \sigma_1 + \vec \sigma_2) \cdot 
(\vec q \times \vec p) [G_0+2G_1]  \nonumber \\ 
&& +4c_4 \vec \tau_1 \cdot \vec \tau_2 \vec \sigma_1 \cdot (\vec q \times \vec p) 
\vec \sigma_2 \cdot (\vec q \times \vec p) [G_0+4G_1+4G_3] \} .
\label{vmed3nl}
\end{eqnarray}
When substituting in the local regulator functions we obtain
\begin{eqnarray}
&&\hspace{-.2in}V_{NN}^{\rm med, 3} = \frac{g_A^2}{16\pi^2 f_\pi^4} \{ -12c_1m_\pi^2\left[ 2\Gamma_0^\prime 
- (2m_\pi^2+q^2) G_0^\prime \right ] - c_3[12\, \Gamma^\prime_4-12(2m_\pi^2+q^2)\Gamma_0^\prime -6q^2\Gamma_1^\prime \nonumber \\
&& + 3(2m_\pi^2+q^2)^2G_0^\prime ] +4c_4 \vec \tau_1 \cdot \vec \tau_2 ( \vec \sigma_1 \cdot \vec \sigma_2 q^2
- \vec \sigma_1 \cdot \vec q \, \vec \sigma_2 \cdot \vec q )G_2^\prime-(3c_3+c_4 \vec \tau_1 \cdot \vec \tau_2 ) 
i (\vec \sigma_1 + \vec \sigma_2) \cdot (\vec q \times \vec p) \nonumber \\
&& [2\Gamma_0^\prime+2\Gamma_1^\prime -(2m_\pi^2 + q^2)(G_0^\prime+2G_1^\prime) ]
 - 12 c_1 m_\pi^2 i (\vec \sigma_1 + \vec \sigma_2) \cdot 
(\vec q \times \vec p) [G_0^\prime+2G_1^\prime] \nonumber \\
&&+4c_4 \vec \tau_1 \cdot \vec \tau_2 \vec \sigma_1 \cdot (\vec q \times \vec p) 
\vec \sigma_2 \cdot (\vec q \times \vec p) [G_0^\prime+4G_1^\prime+4G_3^\prime] \} .
\end{eqnarray}
In the above expressions we encounter the $p$- and $q$-dependent functions
\begin{eqnarray}
&& \hspace{-.9in}G^\prime_{0,*,**}(p,q) = \int_0^{k_f} dk \int_{-1}^1 dx \int_{0}^{2\pi} d\phi\, \frac{\{k^2,k^4,k^6\}/(2\pi)}{A^2-B^2\cos^2 \phi} 
F(p^2 + k^2 + k x \sqrt{4p^2-q^2} \nonumber \\
&& + q k \sqrt{1-x^2} \cos \phi,\Lambda) F(p^2 + k^2 + k x \sqrt{4p^2-q^2} - q k \sqrt{1-x^2} \cos \phi,\Lambda),
\label{gfunlr}
\end{eqnarray}
where $ A = m_\pi^2 + p^2 + k^2 + k x \sqrt{4p^2-q^2} $ and $ B = q k \sqrt{1-x^2} $. In Eq.\ (\ref{vmed3nl}), the functions
$G_{0,*,**}(p,q)$ are obtained from Eq.\ (\ref{gfunlr}) by substituting $\Lambda \rightarrow \infty$. In addition we encounter
the following $p$- and $q$-dependent functions
\begin{equation}
G^\prime_1(p,q) = \frac{\Gamma_0^\prime-(m_\pi^2+p^2)G_0^\prime-G^\prime_*}{4p^2-q^2},
\end{equation}
\begin{equation}
G^\prime_{1*}(p,q) = \frac{3\Gamma_2^\prime+p^2\Gamma_3^\prime-(m_\pi^2+p^2)G^\prime_*-G^\prime_{**}}{4p^2-q^2},
\end{equation}
\begin{equation}
G^\prime_2(p,q) = (m_\pi^2+p^2)G^\prime_1+G^\prime_*+G^\prime_{1*},
\end{equation}
\begin{equation}
G^\prime_3(p,q) = \frac{\frac{1}{2}\Gamma_1^\prime-2(m_\pi^2+p^2)G^\prime_1-2G^\prime_{1*}-G^\prime_*}{4p^2-q^2},
\end{equation}
where
\begin{eqnarray}
&&\hspace{-.6in}\Gamma_0^\prime(p,q) = \int_0^{k_f} dk \int_{-1}^1 dx \int_{0}^{2\pi} d\phi\, 
\frac{k^2/(2\pi)}{m_\pi^2 + p^2 + k^2 + k x \sqrt{4p^2-q^2} + q k \sqrt{1-x^2} \cos \phi} 
\nonumber \\ 
&& \times F(p^2 + k^2 + k x \sqrt{4p^2-q^2} + q k \sqrt{1-x^2} \cos \phi,\Lambda) \nonumber \\
&& \times F(p^2 + k^2 + k x \sqrt{4p^2-q^2} - q k \sqrt{1-x^2} \cos \phi,\Lambda),
\end{eqnarray}
\begin{eqnarray}
&&\hspace{-.6in}\Gamma_1^\prime(p,q) = \int_0^{k_f} dk \int_{-1}^1 dx \int_{0}^{2\pi} d\phi\, 
\frac{k^3[x \sqrt{4p^2-q^2} + q \sqrt{1-x^2}\cos \phi] / (4 \pi p^2)}{m_\pi^2 + p^2 + k^2 + k x \sqrt{4p^2-q^2} + q k \sqrt{1-x^2} \cos \phi} 
 \nonumber \\ 
&& \times F(p^2 + k^2 + k x \sqrt{4p^2-q^2} + q k \sqrt{1-x^2} \cos \phi,\Lambda) \nonumber \\ 
&& \times F(p^2 + k^2 + k x \sqrt{4p^2-q^2} - q k \sqrt{1-x^2} \cos \phi,\Lambda),
\end{eqnarray}
\begin{eqnarray}
&&\hspace{-.6in} \Gamma_2^\prime(p,q) = \int_0^{k_f} dk \int_{-1}^1 dx \int_{0}^{2\pi} d\phi\, 
\frac{k^4\left [4p^2-\left (x\sqrt{4p^2-q^2} + q\sqrt{1-x^2}\cos \phi \right)^2 \right ] 
/ (16 \pi p^2) }{m_\pi^2 + p^2 + k^2 + k x \sqrt{4p^2-q^2} + q k \sqrt{1-x^2} \cos \phi} 
 \nonumber \\
&& \times F(p^2 + k^2 + k x \sqrt{4p^2-q^2} + q k \sqrt{1-x^2} \cos \phi,\Lambda)  \nonumber \\
&& \times F(p^2 + k^2 + k x \sqrt{4p^2-q^2} - q k \sqrt{1-x^2} \cos \phi,\Lambda),
\end{eqnarray}
\begin{eqnarray}
&&\hspace{-.6in} \Gamma_3^\prime(p,q) = \int_0^{k_f} dk \int_{-1}^1 dx \int_{0}^{2\pi} d\phi\, 
\frac{k^4\left [ 3 \left (x\sqrt{4p^2-q^2} + q\sqrt{1-x^2}\cos \phi \right)^2 - 4p^2 \right ] 
/ (16 \pi p^4) }{m_\pi^2 + p^2 + k^2 + k x \sqrt{4p^2-q^2} + q k \sqrt{1-x^2} \cos \phi} 
\nonumber \\
&& \times F(p^2 + k^2 + k x \sqrt{4p^2-q^2} + q k \sqrt{1-x^2} \cos \phi,\Lambda) \nonumber \\
&& \times F(p^2 + k^2 + k x \sqrt{4p^2-q^2} - q k \sqrt{1-x^2} \cos \phi,\Lambda),
\end{eqnarray}
Additionally, we have replaced the quantity $8k_f^3$ in Eq.\ (\ref{vmed3nl}) 
with $12 \Gamma_4^\prime$ defined in Eq.\ (\ref{gam4p}). The term $8k_f^3$ in Eq.\ (\ref{vmed3nl}) as 
well as all unprimed $\Gamma$ and $G$ functions
can be obtained by setting $\Lambda \rightarrow \infty$.
For the $c_D$ vertex correction to one-pion exchange, Fig.\ \ref{fig:mednn}(d), we previously had
\begin{equation}
V_{NN}^{\rm med,4} = -\frac{g_Ac_Dk_f^3}{12\pi^2f_\pi^4\Lambda_\chi} \frac{\vec \sigma_1 \cdot \vec q\,
\vec \sigma_2 \cdot \vec q}{m_\pi^2+q^2}\vec \tau_1 \cdot \vec \tau_2.
\end{equation}
Including the local regulators we find
\begin{equation}
V_{NN}^{\rm med,4} = -\frac{g_Ac_D}{8\pi^2f_\pi^4\Lambda_\chi} \frac{\vec \sigma_1 \cdot \vec q\,
\vec \sigma_2 \cdot \vec q}{m_\pi^2+q^2}\vec \tau_1 \cdot \vec \tau_2 \left ( 
\frac{4}{3}k_f^3 F(q^2,\Lambda) -\Gamma_4^\prime \right ).
\end{equation}
For the $c_D$ vertex correction to the $2N$ contact term, Fig.\ \ref{fig:mednn}(e), we previously had
\begin{eqnarray}
&& V_{NN}^{\rm med,5} = \frac{g_Ac_D}{16\pi^2f_\pi^4\Lambda_\chi} \left \{
\vec \tau_1 \cdot \vec \tau_2 \left [ 2 \vec \sigma_1 \cdot \vec \sigma_2 \Gamma_2 +
\left ( \vec \sigma_1 \cdot \vec \sigma_2 \left ( 2p^2-\frac{q^2}{2} \right ) + \vec \sigma_1 \cdot \vec q\,
\vec \sigma_2 \cdot \vec q \left ( 1-\frac{2p^2}{q^2} \right ) \right . \right . \right . \nonumber \\  
&& \left . \left . \left . - \frac{2}{q^2} \vec \sigma_1 \cdot (\vec q \times \vec p) 
\vec \sigma_2 \cdot (\vec q \times \vec p) \right ) (\Gamma_0+2\Gamma_1+\Gamma_3) \right ] +4k_f^3 - 6m_\pi^2 \Gamma_0\right \}.
\end{eqnarray}
Including the local regulators we obtain
\begin{eqnarray}
&& V_{NN}^{\rm med,5} = \frac{g_Ac_D}{16\pi^2f_\pi^4\Lambda_\chi} \left \{
\vec \tau_1 \cdot \vec \tau_2 \left [ 2 \vec \sigma_1 \cdot \vec \sigma_2 \tilde \Gamma_2 +
\left ( \vec \sigma_1 \cdot \vec \sigma_2 \left ( 2p^2-\frac{q^2}{2} \right ) + \vec \sigma_1 \cdot \vec q\,
\vec \sigma_2 \cdot \vec q \left ( 1-\frac{2p^2}{q^2} \right ) \right . \right . \right . \nonumber \\ 
&& \left . \left . \left . - \frac{2}{q^2} \vec \sigma_1 \cdot (\vec q \times \vec p) 
\vec \sigma_2 \cdot (\vec q \times \vec p) \right ) (\tilde \Gamma_0+2\tilde \Gamma_1+\tilde \Gamma_3) \right ] + 6\tilde \Gamma_4 - 6m_\pi^2 \tilde \Gamma_0\right \} F(q^2,\Lambda).
\end{eqnarray}
We reiterate that the above expressions are obtained in the center-of-mass frame assuming on-shell scattering conditions.
In all cases, the expressions for the in-medium 2N interaction above are well behaved (no poles) and involve only elementary 
integrations.

\section{Applications of density-dependent 2N interactions to nuclear many-body systems}

\subsection{Equation of state of cold nuclear matter}

The equation of state of nuclear matter gives important insights into many properties of finite nuclei, including the 
volume and symmetry energy contributions to the binding energy in the semi-empirical mass formula, the saturated
central density of medium-mass and heavy nuclei, as well as nuclear collective excitation modes and giant resonances. 
The equation of state is also essential for modeling neutron stars 
\cite{lattimer00,steiner05,lattimer07,hebeler10prl,annala18,most18,lim18a,landry18,lim18b}, including their birth in 
core-collapse supernovae, their radii as a function of mass, their tidal deformabilities in the presence of 
compact binary companions, and their moments of inertia. 
For relatively soft equations of state, the central densities of typical neutron stars with mass 
$M \simeq 1.4\,M_\odot$ reach $n \simeq 3n_0$ \cite{lim19epja}, where $n_0=0.16$\,fm$^{-3}$ is the nucleon 
number density in the saturated interior of heavy nuclei. At such densities, three-body forces give a large contribution
to the pressure and are therefore critical for understanding neutron star structure. 

The first-order perturbative contribution (Hartree-Fock approximation) 
to the ground state energy of isospin-symmetric nuclear matter is given by 
\begin{equation}
E^{(1)}_{2N} = \frac{1}{2}\sum_{12} \langle 1 2 | \bar{V}_{2N} | 1 2 \rangle n_1 n_2,
\label{e1nn}
\end{equation}
for the antisymmetrized two-body force $\bar V_{2N}$ and
\begin{equation}
E^{(1)}_{3N} = \frac{1}{6}\sum_{123} \langle 1 2 3 | \bar{V}_{3N} | 1 2 3 \rangle n_1 n_2 n_3,
\label{e1nnn}
\end{equation}
for the antisymmetrized three-body force $\bar V_{3N}$. In the above equations,
$n_i = \theta(k_f - |\vec k_i|)$ is the 
zero-temperature Fermi-Dirac distribution function with Fermi momentum $k_f$, and the sum is taken over the 
momentum, spin, and isospin of the occupied states in the Fermi sea.

\begin{figure}[t]
\begin{center}
\includegraphics[height=7.4cm]{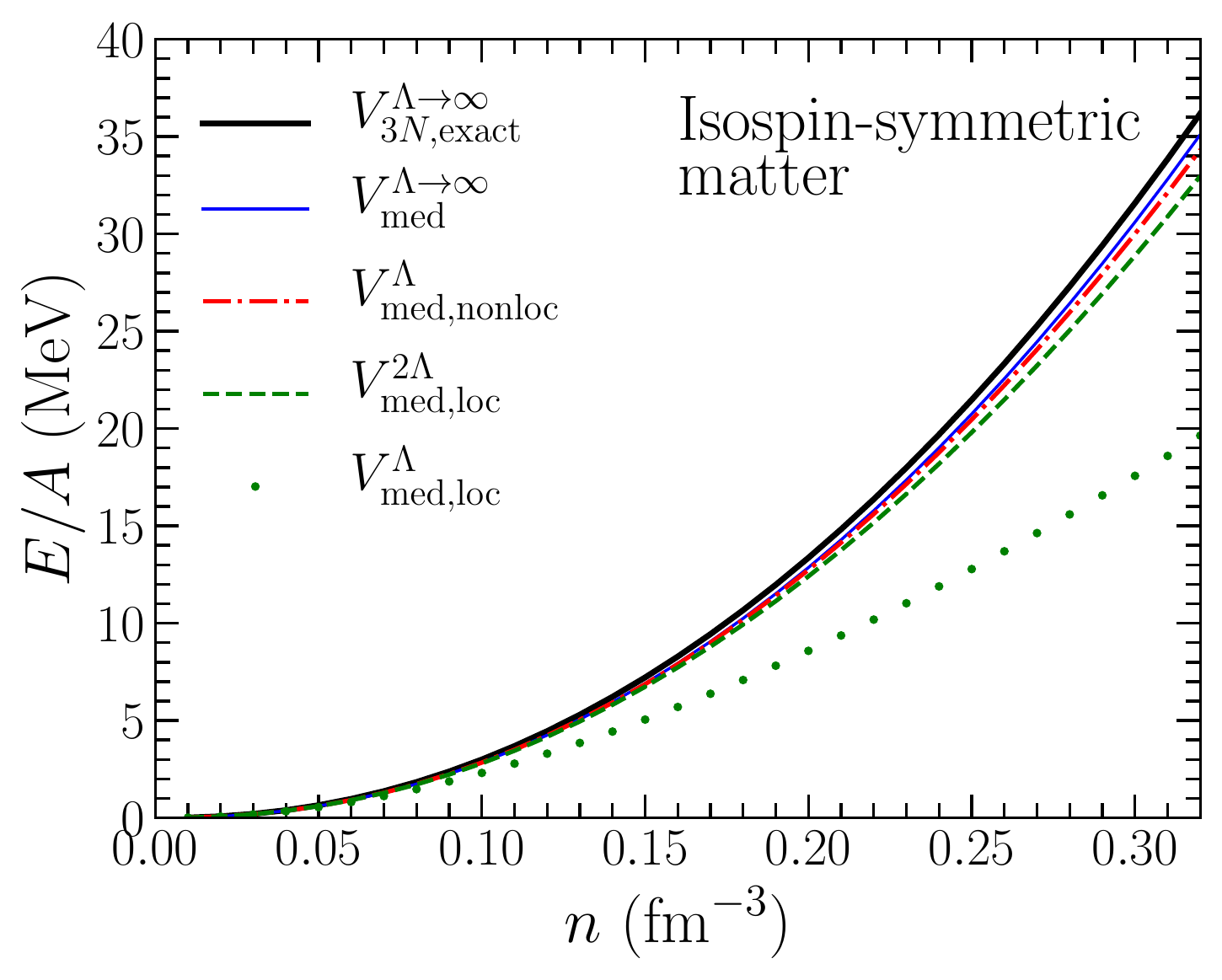}
\end{center}
\caption{Hartree-Fock contribution to the ground-state energy of isospin-symmetric nuclear matter as a function of
density due to the N2LO chiral three-nucleon force with cutoff scale $\Lambda = 450$\,MeV.}
\label{eossnm}
\end{figure}

In Fig.\ \ref{eossnm} we show the density-dependence of the Hartree-Fock contribution to the ground state 
energy of isospin-symmetric nuclear matter from the N2LO chiral three-nucleon force in different approximations. 
As a representative example, we consider the low-energy constants $c_1 = -0.81$\,GeV$^{-1}$, $c_3=-3.4$\,GeV$^{-1}$, 
$c_4=3.4$\,GeV$^{-1}$, $c_D=-0.24$, and $c_E=-0.106$ obtained in Ref.\ \cite{coraggio14} and associated with the N3LO NN
chiral interaction with cutoff scale $\Lambda = 450$\,MeV. We see that in all cases the first-order perturbative contribution
from three-body forces in isospin-symmetric matter is strongly repulsive.
The exact treatment of the Hartree-Fock contribution to the ground state energy arising 
from the N2LO three-nucleon force is shown in Fig.\ \ref{eossnm} as the thick black line labeled 
``$V_{3N,\, \rm exact}^{\Lambda \rightarrow \infty}$''. 
Employing instead the density-dependent NN interaction $V_{\rm med}$ with 
$\Lambda \rightarrow \infty$ we obtain the contribution shown with the thin blue line and labeled 
``$V_{\rm med}^{\Lambda \rightarrow \infty}$''. Note that in order to avoid triple-counting when the density-dependent
NN interaction $V_{\rm med}$ is used in Eq.\ (\ref{e1nn}), we must 
replace $\bar V_{2N} \rightarrow \frac{1}{3}\bar V_{\rm med}$.
From Fig.\ \ref{eossnm} we observe that at the Hartree-Fock level 
the density-dependent NN interaction accurately reflects the physics encoded in the full three-body force. This is
not a trivial observation since several approximations were employed to derive the density-dependent NN interaction
from $V_{3N}$. In Fig.\ \ref{eossnm} we see that the largest
deviation in the two curves is only 1\,MeV (or $\simeq 3\%$) at $n=0.32$\,fm$^{-3}$.

Imposing the nonlocal regulator in Eq.\ (\ref{nlr}) leads to 
the red dot-dashed line labeled ``$V_{\rm med,\, nonloc}^{\Lambda}$''. 
As expected, the presence of the momentum-space cutoff reduces
the Hartree-Fock contribution to the ground-state energy, particularly at high densities. However, the cutoff artifacts
introduced are rather small and amount to only 0.8\,MeV (or $\simeq 2\%$) relative to the result from 
$V_{\rm med}^{\Lambda \rightarrow \infty}$ 
at $n=0.32$\,fm$^{-3}$. We note that since the Hartree-Fock contribution to the ground state energy $E/A$ is 
always finite and probes only the characteristic physical energy scale of the system, the differences between
$V_{\rm med}^{\Lambda \rightarrow \infty}$ and $V_{\rm med,\, nonloc}^{\Lambda}$ are true regulator artifacts.
We next impose the local regulator in Eq.\ (\ref{lr}), which is shown as the dotted green line
in Fig.\ \ref{eossnm} and labeled ``$V_{\rm med,\, loc}^{\Lambda}$''. For this choice of regulator we find severe cutoff artifacts,
even at low densities where one would expect the role of the regulating function to be minimal. 
For example, at $n=0.10$\,fm$^{-3}$, there is a 19\% relative error between 
$V_{\rm med}^{\Lambda \rightarrow \infty}$ and $V_{\rm med,\, loc}^{\Lambda}$. 
From Eq.\ (\ref{lr}) we expect the regulator to introduce corrections at order
$(Q/\Lambda)^4 \sim (k_f/\Lambda)^4$. The Fermi momentum at this density is $k_f = 225$\,MeV, which for the $\Lambda
=450$\,MeV chiral potential implies an
error of $(k_f/\Lambda)^4 \simeq 6\%$. One key difference between the nonlocal and local regulators of Eqs.\ (\ref{nlr})
and (\ref{lr}) is that the relative momentum ranges from $0 < k < k_f$ while the momentum transfer ranges from 
$0 < q < 2k_f$. Therefore, one naturally expects larger cutoff artifacts for the local regulating function in Eq.\ (\ref{lr}).
Indeed, when the value of the momentum-space cutoff is increased to $2\Lambda$, as can be seen from the dashed
green curve of Fig.\ \ref{eossnm}, the results from employing the local regulator are now comparable to those using the 
nonlocal regulator.

\begin{figure}[t]
\begin{center}
\includegraphics[height=7.4cm]{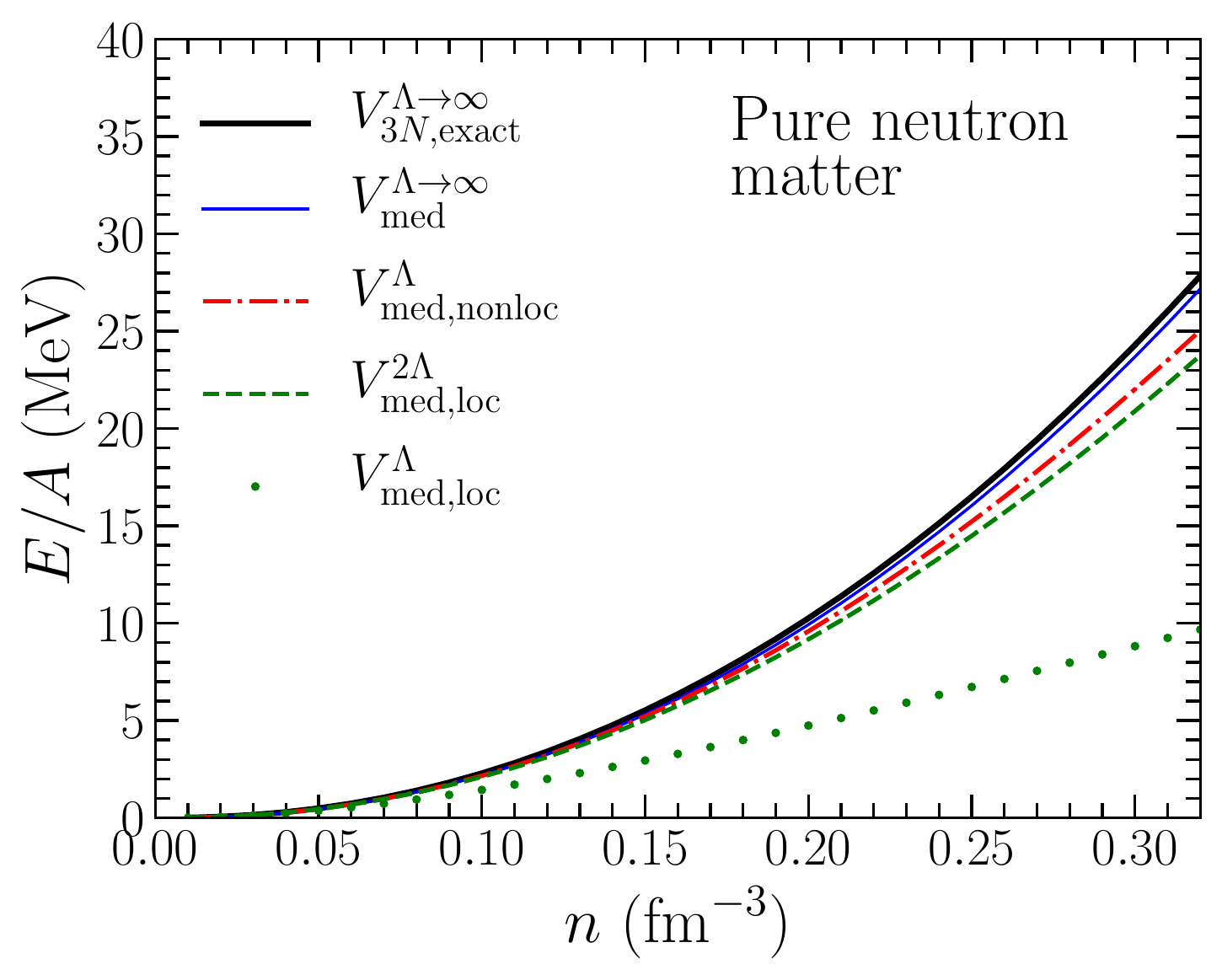}
\end{center}
\caption{Hartree-Fock contribution to the ground-state energy of pure neutron matter as a function of
density due to the N2LO chiral three-nucleon force with cutoff scale $\Lambda = 450$\,MeV.}
\label{eospnm}
\end{figure}

In Fig.\ \ref{eospnm} we show the density-dependence of the Hartree-Fock contribution to the ground state 
energy of pure neutron matter from the N2LO chiral three-nucleon force in different approximations. 
Again we consider the low-energy constants $c_1 = -0.81$\,GeV$^{-1}$, $c_3=-3.4$\,GeV$^{-1}$, 
$c_4=3.4$\,GeV$^{-1}$, $c_D=-0.24$, and $c_E=-0.106$ associated with the N3LO NN
chiral interaction with cutoff scale $\Lambda = 450$\,MeV. However, in pure neutron matter the
Hartree-Fock contribution from three-body forces is independent of $c_4$, $c_D$, and $c_E$. 
We show as the thick black line labeled ``$V_{3N,\, \rm exact}^{\Lambda \rightarrow \infty}$'' 
in Fig.\ \ref{eospnm} the Hartree-Fock contribution to the 
ground-state energy of pure neutron matter. Employing the density-dependent NN interaction $V_{\rm med}$ with 
$\Lambda \rightarrow \infty$ we obtain the contribution shown with the thin blue line labeled 
``$V_{\rm med}^{\Lambda \rightarrow \infty}$''. Again, at the Hartree-Fock level 
the density-dependent NN interaction very accurately reproduces the result from the full three-body force.

Inserting the nonlocal regulator in Eq.\ (\ref{nlr}) we find
the red dot-dashed line labeled ``$V_{\rm med,\, nonloc}^{\Lambda}$''. 
The nonlocal regulator preserves the property that none of the 
three-body force terms proportional to $c_4$, $c_D$, and $c_E$ contribute to the ground-state energy
of pure neutron matter. We find that the momentum-space cutoff reduces
the Hartree-Fock contribution to the ground-state energy even more than that in isospin-symmetric nuclear matter. 
This is due to the larger neutron Fermi momentum (compared to the nucleon Fermi momentum in isospin-symmetric
nuclear matter at the same density).
The cutoff artifacts introduced are nevertheless relatively small and amount to 2\,MeV at $n=0.32$\,fm$^{-3}$. 
Finally, we impose the local regulator in Eq.\ (\ref{lr}) to obtain the dotted green line labeled ``$V_{\rm med,\, loc}^{\Lambda}$''
in Fig.\ \ref{eospnm}. Again, the cutoff artifacts are very large.
For example, at $n=0.10$\,fm$^{-3}$, there is now a 34\% relative error between 
$V_{\rm med}^{\Lambda \rightarrow \infty}$ and $V_{\rm med,\, loc}$. 
At this density, the maximum momentum transfer is $q = 2 k_f \simeq 570$\,MeV, which is clearly problematic
for the chosen cutoff $\Lambda = 450$\,MeV. In addition to larger artifacts, the local regulator also induces contributions
to the density-dependent NN interaction in pure neutron matter that now depend on the low-energy
constants $c_4$, $c_D$ and $c_E$. In the case of pure neutron matter, diagrams ($d$), ($e$), and ($f$) in
Fig.\ \ref{fig:mednn} produce $V_{NN}^{\rm med,4} = V_{NN}^{\rm med,5}
= V_{NN}^{\rm med,6} = 0$ with either the nonlocal regulator or no regulator at all. Instead, for the local
regulator we find
\begin{equation}
V_{NN}^{\rm med,4} = -\frac{g_Ac_D}{8\pi^2f_\pi^4\Lambda_\chi} \frac{\vec \sigma_1 \cdot \vec q\,
\vec \sigma_2 \cdot \vec q}{m_\pi^2+q^2} \left ( 
\frac{2}{3}k_f^3 F(q^2,\Lambda) -\Gamma_4^\prime \right ),
\end{equation}
which indeed vanishes when the local regulators are replaced by $1$.
For the $c_D$ vertex correction to the $2N$ contact term with local regulators we obtain
\begin{eqnarray}
V_{NN}^{\rm med,5} &=& \frac{g_Ac_D}{16\pi^2f_\pi^4\Lambda_\chi} \left \{
 2 \vec \sigma_1 \cdot \vec \sigma_2 \tilde \Gamma_2 +
\left ( \vec \sigma_1 \cdot \vec \sigma_2 \left ( 2p^2-\frac{q^2}{2} \right ) + \vec \sigma_1 \cdot \vec q\,
\vec \sigma_2 \cdot \vec q \left ( 1-\frac{2p^2}{q^2} \right ) \right . \right .\\ \nonumber 
&& \left . \left . - \frac{2}{q^2} \vec \sigma_1 \cdot (\vec q \times \vec p) 
\vec \sigma_2 \cdot (\vec q \times \vec p) \right ) (\tilde \Gamma_0+2\tilde \Gamma_1+\tilde \Gamma_3) + 2\tilde \Gamma_4 - 2m_\pi^2 \tilde \Gamma_0\right \} F(q^2,\Lambda),
\end{eqnarray}
where $\tilde \Gamma_4(p)$ is defined in Eq.\ (\ref{gam4t}).
Finally, the three-body contact term with the local regulator leads to
\begin{equation}
V_{NN}^{\rm med, 6} = \frac{c_E}{2\pi^2 f_\pi^4 \Lambda_\chi} \left [\frac{2}{3}k_f^3\,
F^2(q^2,\Lambda) -2 F(q^2,\Lambda) \tilde \Gamma_4(p) - \Gamma^\prime_4(p,q) + \frac{4}{3}k_f^3 F(q^2,\Lambda)
\right ].
\end{equation}
Again, this term vanishes when the regulating functions are set to $1$. These additional terms have been 
included in the present of calculation of the dotted green line in Fig.\ \ref{eospnm}.
Substituting $\Lambda \rightarrow 2 \Lambda$ into the nonlocal regulator
again reduces the cutoff artifacts, as seen in the dashed green curve of Fig.\ \ref{eospnm}.

The inclusion of three-body forces in the nuclear equation of state beyond the Hartree-Fock approximation remains
challenging. While several recent works \cite{kaiser12,drischler19} have computed the exact second-order contribution 
to the equation of state from three-body forces, the use of derived density-dependent two-body interactions allows for an
approximate treatment up to third order in perturbation theory \cite{coraggio14,holt17prc}:
\begin{equation}
E^{(2)} = -\frac{1}{4} 
\sum_{1234} \left| \langle 1 2 \left | \bar{V}_{\text{eff}} \right | 3 4 \rangle \right |^2
\frac{n_1 n_2 \bar{n}_3 \bar{n}_4}
{e_3+e_4-e_1-e_2},
\label{e2}
\end{equation}
\begin{equation}
E^{(3)}_{pp} = \frac{1}{8} 
\sum_{123456} \langle 1 2 \left | \bar{V}_{\text{eff}} \right | 3 4 \rangle
\langle 3 4 \left | \bar{V}_{\text{eff}} \right | 5 6 \rangle
\langle 5 6 \left | \bar{V}_{\text{eff}} \right |  1 2 \rangle
 \frac{n_1 n_2 \bar{n}_3 \bar{n}_4 \bar{n}_5 \bar{n}_6}
{(e_3+e_4-e_1-e_2)(e_5+e_6-e_1-e_2)},
\label{e3pp}
\end{equation}
\begin{equation}
E^{(3)}_{hh} = \frac{1}{8} 
\sum_{123456} \langle 1 2 \left | \bar{V}_{\text{eff}} \right | 3 4 \rangle
\langle 3 4 \left | \bar{V}_{\text{eff}} \right | 5 6 \rangle
\langle 5 6 \left | \bar{V}_{\text{eff}} \right |  1 2 \rangle
\frac{\bar{n}_1 \bar{n}_2 n_3 n_4 n_5 n_6}
{(e_1+e_2-e_3-e_4)(e_1+e_2-e_5-e_6)},
\label{e3hh}
\end{equation}
\begin{equation}
E^{(3)}_{ph} = 
-\sum_{123456} \langle 1 2 \left | \bar{V}_{\text{eff}} \right | 3 4 \rangle
\langle 5 4 \left | \bar{V}_{\text{eff}} \right | 1 6 \rangle
\langle 3 6 \left | \bar{V}_{\text{eff}} \right | 5 2 \rangle
\frac{n_1 n_2 \bar{n}_3 \bar{n}_4 n_5 \bar{n}_6}
{(e_3+e_4-e_1-e_2)(e_3+e_6-e_2-e_5)},
\label{e3ph}
\end{equation}
where $\bar n_j=1-n_j$ and $V_{\text{eff}} = V_{2N}+ V_{\rm med}$. The intermediate-state single-particle energies $e_i$
in Eqs.\ (\ref{e2})$-$(\ref{e3ph}) can be treated in several different approximations. In the simplest case, they are taken as
the free-space energies: $e(k) = k^2/2M$. More generally, they can be dressed with interaction lines \cite{holt13prc} in which
case $e(k) = k^2/2M + \Re \Sigma(e(k),k)$, where $\Sigma(e(k),k)$ is the self-consistent energy- and momentum-dependent 
nucleon self energy.

\begin{figure}[t]
\begin{center}
\includegraphics[height=8.6cm]{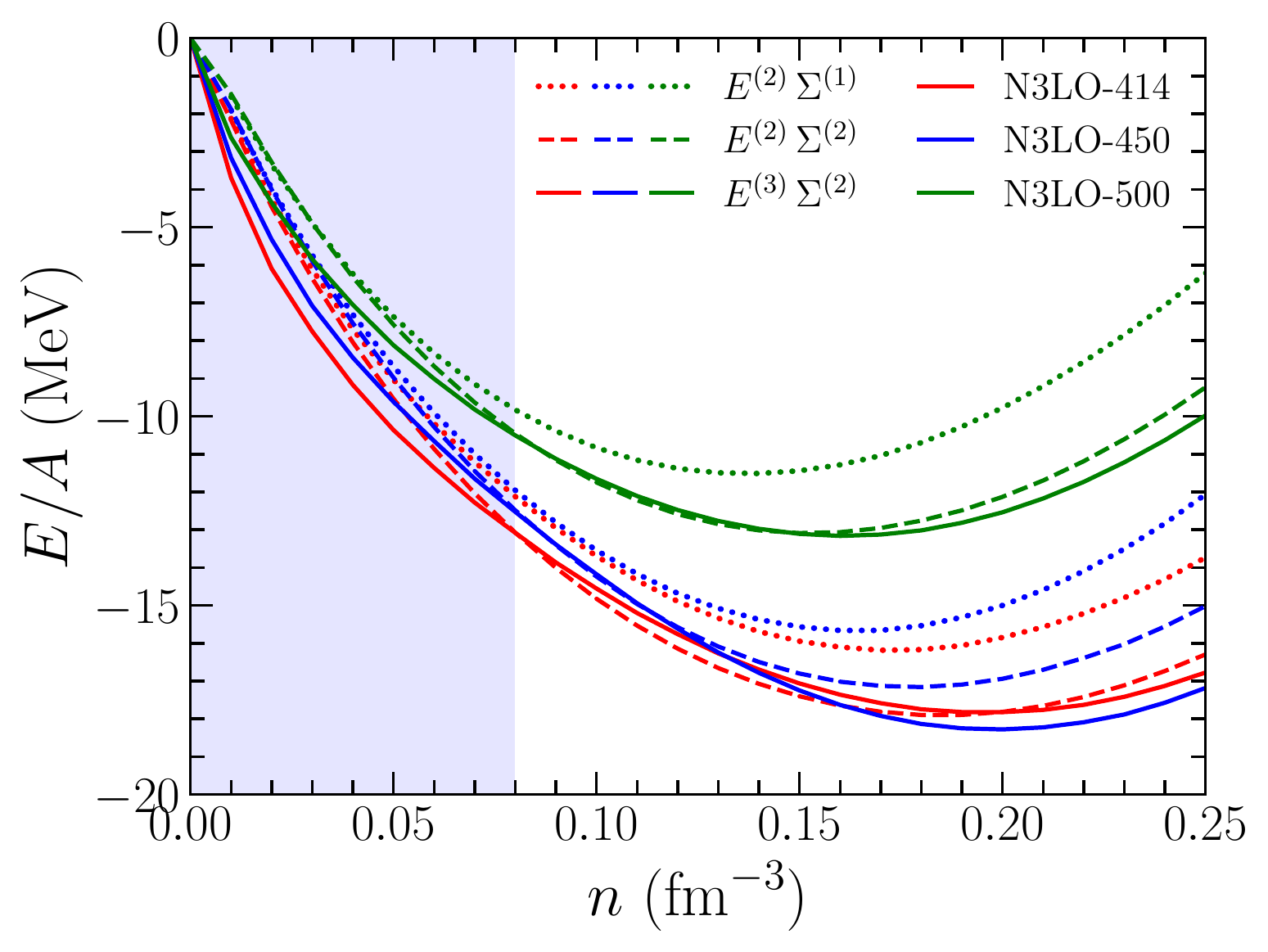}
\end{center}
\caption{Equation of state of isospin-symmetric nuclear matter from chiral two- and three-body forces with different
choices of the momentum-space cutoff $\Lambda$ and at different orders in many-body perturbation theory. The 
label $E^{(i)}$ denotes the $i$-th order in perturbation theory, and $\Sigma^{(n)}$ denotes the $n$-th order treatment of
the self-energy. The shaded region below $n=0.08$\,fm$^{-3}$ represents the approximate location of the 
spinodal instability.}
\label{eos3ph}
\end{figure}

Third-order diagrams
\cite{coraggio14} and fourth-order diagrams \cite{drischler19} are found to give rather small contributions ($\sim 2$\,MeV)
to the 
equation of state up to $n=1.5 n_0$ for potentials with momentum-space cutoffs $\Lambda \simeq 400 - 500$\,MeV. 
However, the intermediate-state energies in Eqs.\ (\ref{e2})$-$(\ref{e3ph}) should be treated at least to second order 
\cite{holt17prc} in a perturbative expansion of the self-energy. In Fig.\ \ref{eos3ph} we plot the equation of state
of isospin-symmetric nuclear matter for several different choices of the cutoff scale $\Lambda = 414, 450, 500$\,MeV
(represented by red, blue, and green colors respectively) and orders in many-body perturbation theory (denoted by the 
symbol). In all cases we employ an N3LO chiral nucleon-nucleon interaction with only the N2LO chiral three-body force
with low-energy constants fitted to the binding energies of $^3$H and $^3$He as well as the beta-decay lifetime of
$^3$H. For the density-dependent three-body force we use the nonlocal regulator in Eq.\ (\ref{nlr}). 
From Figs.\ \ref{eossnm} and \ref{eospnm} we see that the local regulator in Eq.\ (\ref{lr}) 
would be highly constraining and only allow for a meaningful calculation of the nuclear equation of state below saturation
density. In Fig.\ \ref{eos3ph}
the dotted lines denote the inclusion of second-order ground-state energy diagrams ($E^{(2)}$) with first-order 
self energies ($\Sigma^{(1)}$) for the intermediate-state propagators. The dashed lines denote the inclusion of 
second-order ground-state energy diagrams ($E^{(2)}$) with second-order self energies ($\Sigma^{(2)}$) for the 
intermediate-state propagators. From Fig.\ \ref{eos3ph} we see that the second-order self energy diagrams contribute
$2-3$\,MeV to the ground state energy per particle for densities $n \ge 0.16$\,fm$^{-3}$. Finally, the solid lines denote 
the inclusion of third-order ground-state energy diagrams ($E^{(3)}$) with second-order self energies ($\Sigma^{(2)}$) for the 
intermediate-state propagators. In general, the sum of all third-order diagrams gives a relatively small contribution to
the equation of state around saturation density. However, below the critical density for the spinodal instability ($n_c \simeq
0.08$\,fm$^{-3}$) \cite{wellenhofer15}, denoted by the blue shaded region in Fig.\ \ref{eos3ph},
the third-order diagrams give somewhat large effects due to the breakdown of perturbation theory.
Nevertheless, the saturation of nuclear matter is robust and both the empirical saturation density and energy are
within the uncertainties predicted from chiral nuclear forces. We note that the ground state energy from the N3LO-414 and
N3LO-450 chiral potentials are very similar in all approximations. Both potentials are known to converge very rapidly
in perturbation theory compared to the N3LO-500 potential \cite{holt17prc}, which may partly explain the similarity of
their results.

\subsection{Nucleon-nucleus optical potentials}

The theoretical description of nucleon-nucleus scattering and reactions can be greatly simplified through
the introduction of optical model potentials, which replace the complicated two- and many-body interactions between
projectile and target with an average one-body potential. In many-body perturbation theory, the optical 
potential can be identified as the nucleon self-energy, which in general is complex, nonlocal, and energy dependent:
\begin{equation}
V(\vec r, \vec r^{\, \prime};E) = U(\vec r, \vec r^{\, \prime};E) + i W(\vec r, \vec r^{\ \prime};E).
\label{omp}
\end{equation}
While phenomenological optical potentials \cite{koning03} are fitted to a great amount of differential elastic scattering, 
total cross section, and analyzing power data, microscopic optical potentials can be constructed from high-precision 
two-nucleon and three-nucleon forces \cite{jeukenne77,Toyokawa3N,vorabbi16,Rotureau18,idini19}. 
In chiral effective field theory, three-nucleon forces in particular have been shown \cite{holt13prc,holt16prc} to give rise to an
overall repulsive single-particle potential at all projectile energies that increases strongly with the density of the
medium. Three-nucleon forces are therefore essential for an accurate description of nucleon-nucleus scattering
at moderate energies where the projectile penetrates the target nucleus.
\begin{figure}[t]
\begin{center}
\includegraphics[height=8cm]{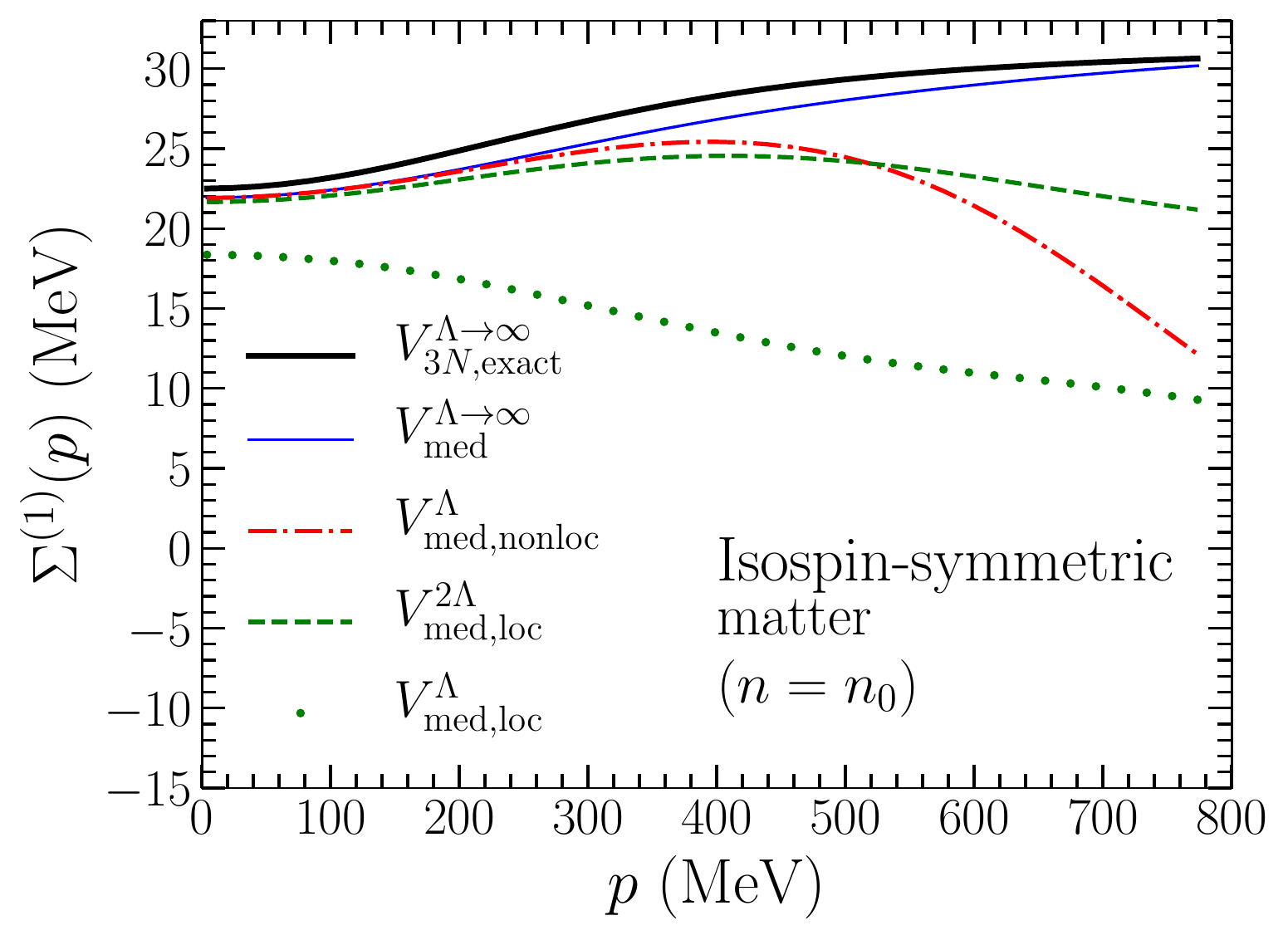}
\end{center}
\caption{Hartree-Fock contribution to the nucleon self energy in symmetric nuclear matter at saturation density
$n_0$ from the N2LO chiral three-nucleon force with cutoff scale $\Lambda = 450$\,MeV. Results are shown
for an exact treatment as well as from the density-dependent interaction $V_{\rm med}$ with different choices
of regulating function (see text).}
\label{spesnm}
\end{figure}

In the Hartree-Fock approximation, the contribution to the nonlocal (but energy-independent) 
nucleon self energy is given by
\begin{equation}
\Sigma^{(1)}_{2N}(q) = \sum_{1} \langle \vec q \, \vec h_1 s s_1 t t_1 | \bar V_{2N} | \vec q \,
\vec h_1 s s_1 t t_1 \rangle n_1,
\label{se1}
\end{equation}
where $\bar V_{2N}$ denotes the antisymmetrized NN potential, $n_1 = \theta(k_f-|\vec h_1|)$ is the 
zero-temperature Fermi-Dirac distribution function, and the sum is taken over the 
momentum, spin, and isospin of the intermediate hole state $|\vec h_1, s_1, t_1\rangle$.
The Hartree-Fock contribution from three-body forces is given by
\begin{equation}
\Sigma^{(1)}_{3N}(q) = \frac{1}{2}\sum_{12} \langle \vec q \, \vec h_1\vec h_2; s s_1s_2; t t_1t_2 |
 \bar V_{3N} | \vec q \, \vec h_1\vec h_2; s s_1s_2; t t_1t_2 \rangle n_1 n_2,
\label{se31}
\end{equation}
where $\bar V_{3N}$ is the fully-antisymmetrized three-body interaction. We have computed the Hartree-Fock
contribution to the single-particle energy exactly \cite{holt13prc} from Eq.\ (\ref{se31}) as well as from Eq.\ (\ref{se1}) 
using the density-dependent NN interaction $V_{\rm med}$. Note that in order to avoid double-counting we must 
replace $\bar V_{2N} \rightarrow \frac{1}{2}\bar V_{\rm med}$ in Eq.\ (\ref{se1}).

In Fig.\ \ref{spesnm} we demonstrate the accuracy of using the density-dependent NN interaction in place of the 
full three-body force when computing the Hartree-Fock contribution to the nucleon self energy. 
Specifically, we plot the momentum-dependent nucleon self-energy (note that both the 2N and 3N 
Hartree-Fock contributions are real and energy independent) in isospin-symmetric nuclear
matter at saturation density $n_0$. The thick black curve labeled ``$V_{3N,\,\rm exact}$'' is the exact result without a 
high-momentum regulator. The thin blue curve labeled ``$V_{\rm med}^{\Lambda \rightarrow \infty}$'' is obtained 
from the density-dependent NN interaction without regulator. We see that there is a systematic difference of $1-2$\,MeV
(or about 5\%) between the two results across all momenta. This difference represents the inherent error introduced 
through the approximations employed in constructing the density-dependent NN interaction. Except for this 
systematic reduction in the nucleon self energy, we see that overall
$V_{\rm med}$ faithfully reproduces the exact Hartree-Fock self-energy across all momenta. 

\begin{figure}[t]
\begin{center}
\includegraphics[height=8cm]{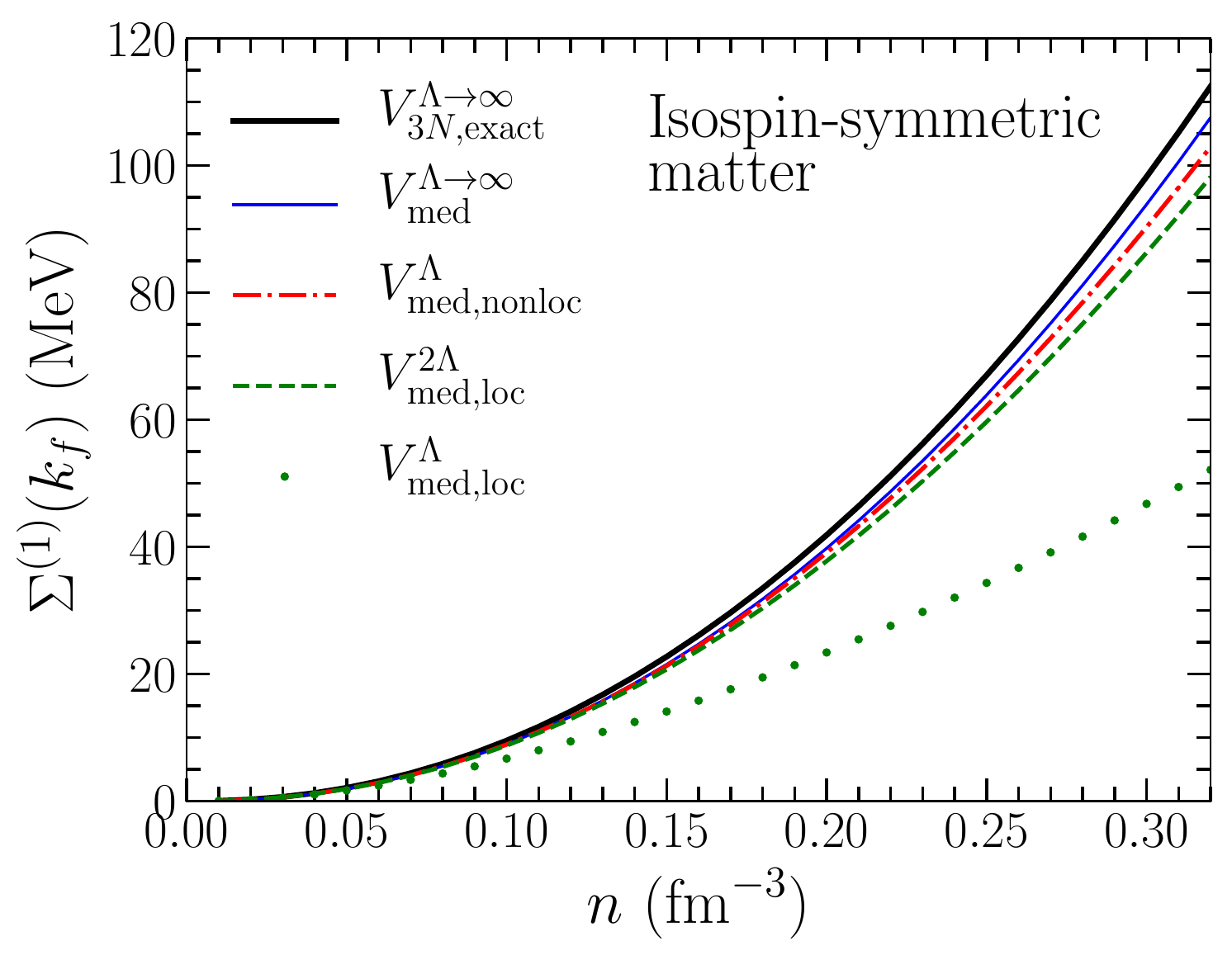}
\end{center}
\caption{Hartree-Fock contribution to the nucleon self energy at the Fermi momentum ($p=k_f$)
in symmetric nuclear matter as a function of density 
from the N2LO chiral three-nucleon force with cutoff scale $\Lambda = 450$\,MeV. Results are shown
for an exact treatment as well as from the density-dependent interaction $V_{\rm med}$ with different choices
of regulating function (see text).}
\label{speqkfsnm}
\end{figure}

Introducing the nonlocal
regulator in Eq.\ (\ref{nlr}) results in the red dash-dotted line of Fig.\ \ref{spesnm}. The artifacts associated
with the nonlocal regulator grow rapidly for momenta beyond $p \simeq 400$\,MeV and by $p\simeq 600$\,MeV
the three-nucleon force contribution to the self energy is reduced by $\sim 25\%$. This corresponds to a lab energy of about 
$E_{\rm lab} \simeq 175$\,MeV \cite{holt16prc} beyond which a description of nucleon-nucleus scattering
in terms of chiral optical potentials becomes highly questionable. Introducing the local regulator in Eq.\ (\ref{lr})
leads to the dotted green curve in Fig.\ \ref{spesnm}. We see that this regulator generates artifacts (of
at least $15\%$) even for low-momentum particles in isospin-symmetric nuclear matter at saturation density.
This is due to the already large nucleon Fermi momentum ($k_f \simeq 270$\,MeV) in nuclear matter 
at this density. Finally, if we double the value of the momentum-space cutoff in the local regulating function, 
we find the results given by the 
dashed green curve in Fig.\ \ref{spesnm}. Again, this choice of cutoff leads to artifacts that are on par with those
from the nonlocal regulator but which are noticeably smaller at the largest momenta considered.

In Fig.\ \ref{speqkfsnm} we plot the value of the Hartree-Fock single-particle potential at the Fermi momentum ($p=k_f$) 
from chiral three-body forces for densities up to $n\simeq 2n_0$. This contribution to the single-particle energy 
from the chiral 3NF grows approximately quadratically with the density. Again we find that the density-dependent 
NN interaction from the leading chiral three-nucleon force reproduces well the exact Hartree-Fock result. The
artifacts introduced through the nonlocal regulator in Eq.\ (\ref{nlr}), the local regulator in Eq.\ (\ref{lr}), and the 
local regulator with $\Lambda_{\rm loc} = 2\Lambda_{\rm nonloc}$ 
follow the same trends already observed in the Hartree-Fock
contribution to the equation of state.

Recently, several works \cite{holt13prc,holt16prc} have included the second-order contributions to the
nucleon self energy (both in isospin-symmetric and asymmetric nuclear matter):
\begin{equation}
\Sigma^{(2a)}_{2N}(q,\omega)
= \frac{1}{2}\sum_{123} \frac{| \langle \vec p_1 \vec p_3 s_1 s_3 t_1 
t_3 | \bar V_{\text{eff}} | \vec q \, \vec h_2 s s_2 t t_2 \rangle |^2}{\omega + e_2 - e_1
-e_3 + i \eta} \bar n_1 n_2 \bar n_3
\label{op2ac}
\end{equation}
and
\begin{equation}
\Sigma^{(2b)}_{2N}(q,\omega)
= \frac{1}{2}\sum_{123} \frac{| \langle \vec h_1 \vec h_3 s_1 s_3 t_1 
t_3 | \bar V_{\text{eff}} | \vec q \, \vec p_2 s s_2 t t_2 \rangle |^2}{\omega + e_2 - e_1
- e_3 - i \eta} n_1 \bar n_2 n_3,
\label{op2bd}
\end{equation}
with the antisymmetrized potential $\bar V_{\text{eff}} = \bar V_{2N} + \bar V_{\rm med}$ that includes the density-dependent 
interaction from the N2LO chiral three-body force. The single-particle energies in Eqs.\ (\ref{op2ac}) and (\ref{op2bd}) 
are computed self-consistently according to
\begin{equation}
\label{eq:1}
e(q)=\frac{q^2}{2M}+\Re \Sigma(e(q),q).
\end{equation}
Generically, Eqs.\ (\ref{op2ac}) and (\ref{op2bd})
give rise to complex and energy-dependent single-particle potentials. This allows for the construction 
of nucleon-nucleus optical potentials that have been shown \cite{whitehead19} to reproduce well differential
elastic scattering cross sections for proton projectiles on a range of calcium targets up to about $E = 150$\,MeV. 

The general form of phenomenological optical potentials for nucleon-nucleus scattering is given by
\begin{equation}
U(r,E) = V_V(r,E) + i W_V(r,E) + i W_D(r,E) 
+ V_{SO}(r,E) \vec \ell \cdot \vec s + i W_{SO}(r,E) \vec \ell \cdot \vec s + V_C(r),
\label{phen}
\end{equation}
consisting of a real volume term, an imaginary volume term, an imaginary surface term, a real spin-orbit term, an 
imaginary spin-orbit term, and finally a central Coulomb interaction. In Eq.\ (\ref{phen}), $\vec \ell$ and $\vec s$ are 
the single-particle orbital angular momentum and spin angular momentum, respectively.
To construct a microscopic nucleon-nucleus optical potential from the nuclear matter approach, one can employ
the local density approximation (LDA):
\begin{equation}
V(E;r) + i W(E;r) = V(E;k_f^p(r),k_f^n(r)) + i W(E;k_f^p(r),k_f^n(r)),
\end{equation}
where $k_f^p(r)$ and $k_f^n(r)$ are the local proton and neutron Fermi momenta. This approach can be improved
by taking account of the finite range of the nuclear force through the improved local density approximation (ILDA):
\begin{equation}
{V}(E;r)_{ILDA}=\frac{1}{(t\sqrt{\pi})^3}\int V(E;r') e^{\frac{-|\vec{r}-\vec{r}'|^2}{t^2}} d^3r',
\label{eq:ilda}
\end{equation}
which introduces an adjustable length scale $t$ taken to be the typical range of the nuclear force. In previous works
\cite{bauge98,whitehead19} this Guassian smearing factor was chosen to be $t\simeq 1.2$\,fm and varied in order to 
estimate the introduced theoretical uncertainties.

\begin{figure}[t]
\begin{center}
\includegraphics[height=12cm]{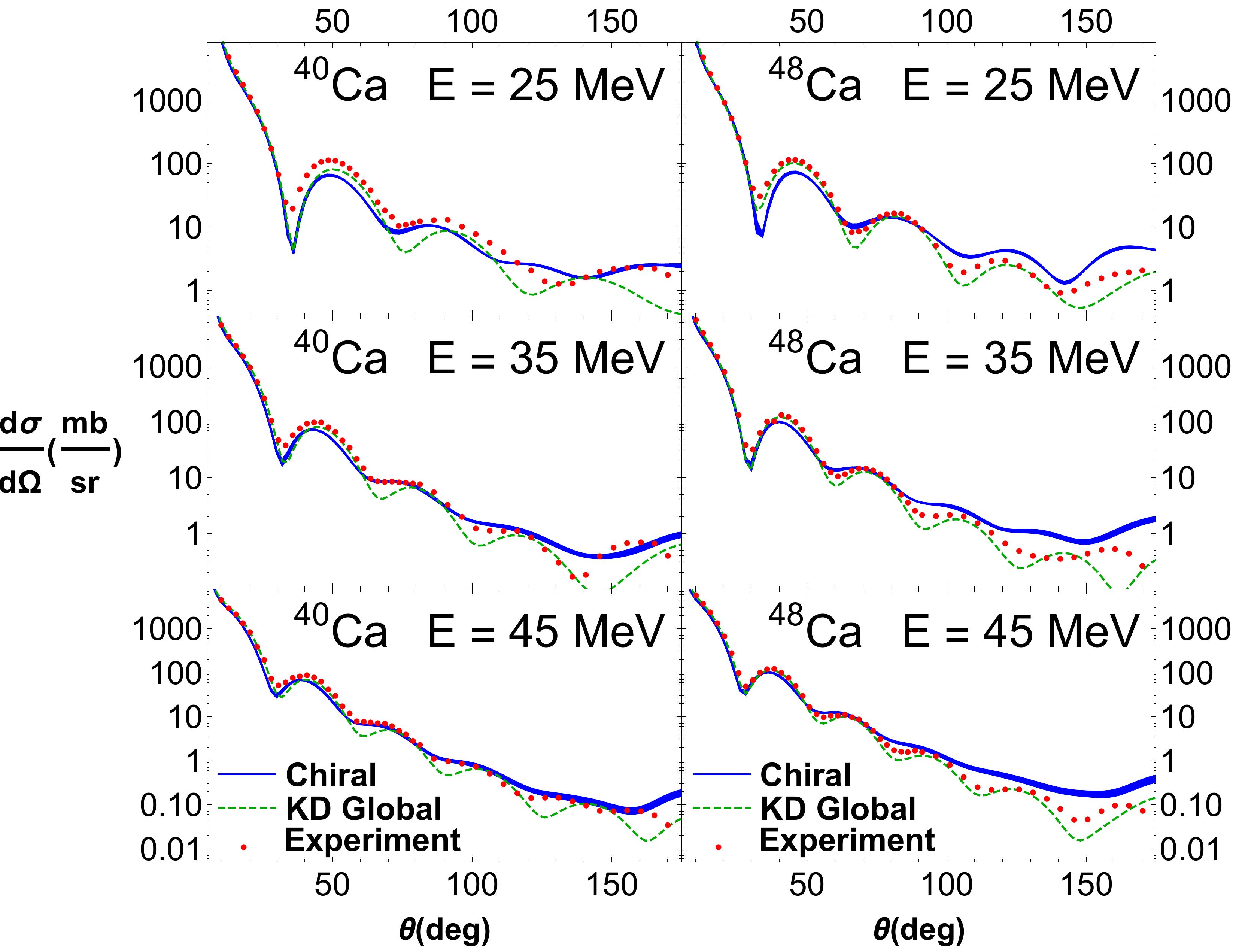}
\end{center}
\caption{Differential elastic scattering cross sections for proton projectiles on $^{40}$Ca and $^{48}$Ca targets 
at the energies $E=25,35,45$\,MeV. The cross sections computed from microscopic chiral optical potentials
including two- and three-body forces are shown as the blue band. The cross sections from the Koning-Delaroche
``KD'' phenomenological optical potential are given by the green dashed curves, and experimental data are 
shown by red circles.}
\label{cagrid}
\end{figure}

The ILDA approach starts by defining the isoscalar and isovector density distributions for a given target nucleus. 
In our previous works \cite{whitehead19,whitehead20}, we have employed for this purpose Skyrme energy 
density functionals fitted to the equation of state of isospin-asymmetric
nuclear matter \cite{lim17} calculated from the same chiral two- and three-body forces used to compute the nucleon self energy
in Eqs.\ (\ref{se1})$-$(\ref{op2bd}). The Gaussian smearing factor $t$ in the ILDA was chosen in the range
$1.15\,{\rm fm} \leq t \leq 1.25\,{\rm fm}$. 
The real part of the optical potential is found \cite{whitehead19} to be in excellent agreement with that from
phenomenological optical potentials
\cite{koning03}, however, the microscopic imaginary part exhibits a surface peak that is too small and a volume 
contribution that grows too strongly with energy. This leads to larger
total reaction cross sections \cite{whitehead19} compared to phenomenology and experiment. This is in fact a general feature
of the microscopic nuclear matter approach \cite{lagrange82,kohno84} independent of the choice of nuclear potential,
and previous works \cite{bauge98,goriely07} have attempted to mitigate this deficiency by introducing scaling factors for the imaginary part.

In Fig.\ \ref{cagrid} we plot the differential elastic scattering cross sections for proton projectiles on $^{40}$Ca and $^{48}$Ca
isotopes from microscopic optical potentials derived in chiral effective field theory. In this study we employ the N3LO 
nucleon-nucleon potential with momentum-space cutoff $\Lambda = 450$\,MeV together with the density-dependent
NN interaction using the nonlocal regulator in Eq.\ (\ref{nlr}). 
From Fig.\ \ref{cagrid} we see that the predictions from chiral effective field theory (shown
in blue) reproduce well the elastic scattering cross section data (red dots) from $E=25-45$\,MeV. The small 
uncertainty band associated with the blue curve is due entirely to variations in the ILDA Gaussian smearing factor.
In some cases,
the results from chiral nuclear optical potentials give better agreement with experiment than the Koning-Delaroche
phenomenological optical potential (shown as the green dashed line in Fig.\ \ref{cagrid}). 
In contrast to semi-microscopic approaches \cite{bauge98,goriely07} that introduce energy-dependent scaling 
factors for the real and imaginary parts of the optical potential, our calculations are not fitted in any way to scattering data.
Qualitatively similar results have been found \cite{whitehead19} for proton energies as low as $E \simeq 2$\,MeV 
and as high as $E \simeq 160$\,MeV. 
Moreover, the construction of {\it neutron}-nucleus optical potentials is in progress \cite{whitehead20} and preliminary results 
for differential elastic scattering cross sections are found to 
be of similar quality to the case of proton-nucleus scattering.

\subsection{Quasiparticle interaction in nuclear matter}

Landau's theory of normal Fermi liquids \cite{landau57a, landau59,baym91} remains a valuable theoretical framework for 
understanding the excitations, response, and transport coefficients of nuclear many-body 
systems \cite{migdal64,migdal67}. Fermi liquid theory is based on the concept of quasiparticles, i.e., dressed 
single-particle excitations of a (potentially) strongly-interacting many-body system that retain key properties of the bare 
particles in the analogous non-interacting system. In this way, Fermi liquid theory allows for
a convenient description of the low-energy excitations of the interacting system and
in the context of the nuclear many-body problem helps justify the nuclear
shell model and the independent-particle description of nuclei and nuclear matter. The theory is made quantitative
through the introduction of the quasiparticle interaction $\mathcal F$, defined as the second functional derivative of 
the energy with respect to the quasiparticle distribution function $n(\vec p\,)$:
\begin{equation}
E = E_0 + \sum_{1} e_{\vec p_1}\, \delta 
n_{\vec p_1 s_1 t_1}
+ \frac{1}{2\Omega} \sum_{1 2}{\mathcal F}({\vec p}_1 s_1 t_1;
{\vec p}_2 s_2 t_2) \delta n_{{\vec p}_1 s_1 t_1}
\delta n_{{\vec p}_2 s_2 t_2},
\label{deltae}
\end{equation}
where $E_0$ is the ground state energy, $\Omega$ is a normalization volume, and 
$\delta n_{\vec p_i s_i t_i}$ is the change in occupation number of state $i$. In Eq.\ (\ref{deltae}) the quasiparticle 
interaction ${\mathcal F}$ in momentum space
has units fm$^2$, $s_i$ labels the spin quantum number of quasiparticle $i$,
and $t_i$ labels the isospin quantum number. Enforcing the symmetries of the strong interaction and
assuming that the quasiparticles lie exactly on the Fermi surface leads
to the general form of the quasiparticle interaction:
\begin{equation}
{\mathcal F}(\vec p_1, \vec p_2\,) = {\mathcal A }(\vec p_1, \vec p_2\,) + 
{\mathcal A^\prime }(\vec p_1, \vec p_2\,) \vec \tau_1 \cdot \vec \tau_2,
\label{qpi1}
\end{equation}
where \cite{schwenk04}
\begin{eqnarray}
&&\hspace{-.3in}{\mathcal A}(\vec p_1, \vec p_2\,) = f(\vec p_1, \vec p_2\,) + g(\vec p_1, 
\vec p_2\,) \vec \sigma_1 \cdot \vec \sigma_2
+ h (\vec p_1, \vec p_2\,) 
S_{12}(\hat p) + k (\vec p_1, \vec p_2\,) S_{12}(\hat P) \nonumber \\
&&+\ell (\vec p_1, \vec p_2\,) (\vec \sigma_1 \times \vec \sigma_2)\cdot 
(\hat p \times \hat P),
\label{qpi}
\end{eqnarray}
and likewise for ${\mathcal A^\prime}$ except with the replacement 
$\{f,g,h,k,\ell\} \longrightarrow \{f^\prime,g^\prime,h^\prime,k^\prime,\ell^\prime \}$.
The relative momentum is given by $\vec p = \vec p_1 - \vec p_2$,
the center-of-mass momentum is defined by $\vec P = \vec p_1 + \vec p_2$, and
the tensor operator has the form
$S_{12}(\hat v) = 3 \vec \sigma_1 \cdot \hat v\, \vec \sigma_2 \cdot \hat 
v -\vec \sigma_1 \cdot\vec \sigma_2$. 

For two quasiparticle momenta on the Fermi surface ($|\vec p_1| = |\vec p_2| =k_f$),
the scalar functions $\{f,g,h,k,\ell,f^\prime,g^\prime,h^\prime,k^\prime,\ell^\prime \}$
depend only the angle $\theta$ between and $\vec p_1$ and $\vec p_2$.
The quasiparticle interaction can therefore be written in terms of Legendre polynomials:
\begin{equation}
f({\vec p}_1,{\vec p}_2) = \sum_{L=0}^\infty f_L(k_f) P_L(\mbox{cos } \theta),\hspace{.2in}
f^\prime({\vec p}_1,{\vec p}_2) = \sum_{L=0}^\infty f^\prime_L(k_f) P_L(\mbox{cos } \theta), \hspace{.2in} \dots
\label{gflp}
\end{equation}
where $\cos \theta ={\hat p}_1 \cdot {\hat p}_2$, $q = 2k_f\, {\rm sin}\, (\theta /2)$,
and $P=2k_f \cos(\theta/2)$.
The coefficients $f_L, f^\prime_L,\dots$ are referred to as the Fermi liquid
parameters. Dimensionless Fermi liquid parameters $F_L, F^\prime_L,\dots$ can be defined by 
multiplying $f_L, f^\prime_L,\dots$ by the density of states, e.g., for symmetric nuclear matter:
\begin{equation}
N_0 = 2 M^* k_f / \pi^2,
\label{dos}
\end{equation}
where $M^*$ the effective nucleon mass.

\begin{figure}[t]
\begin{center}
\includegraphics[height=3.5cm]{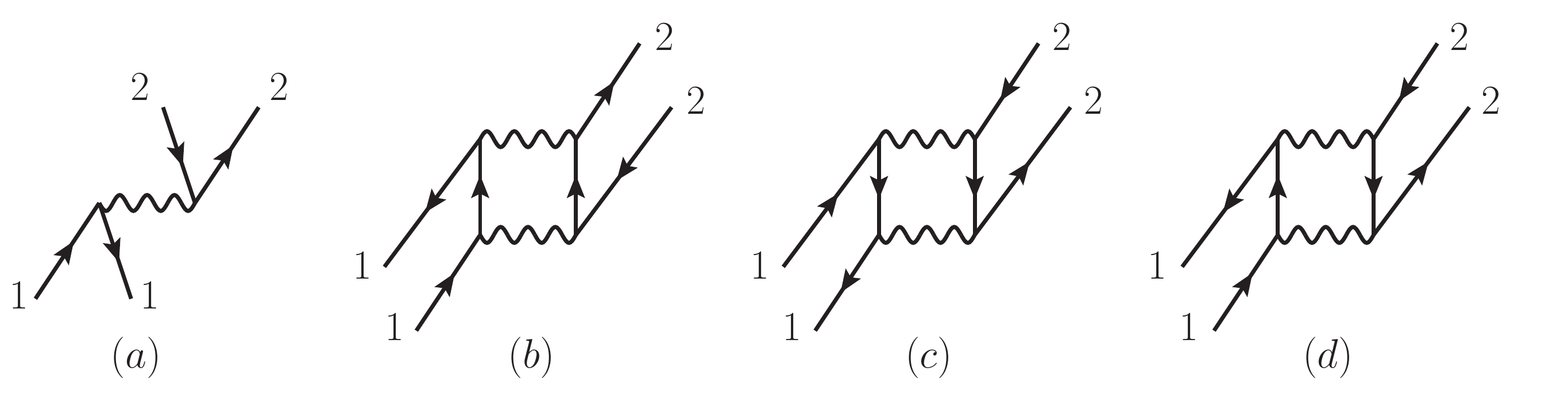}
\end{center}
\vspace{-.5cm}
\caption{Diagrammatic contributions to the quasiparticle interaction up to 
second order in perturbation theory. Wavy lines denote the antisymmetrized nuclear
interaction.}
\label{diag}
\end{figure}

Originally, Fermi liquid theory was treated as a phenomenological model \cite{migdal64} in which the 
lowest-order Fermi liquid parameters would be constrained by select experimental data. From the 
Brueckner-Goldstone linked diagram expansion for the ground-state energy, see e.g., 
Eqs.\ (\ref{e1nn})$-$(\ref{e3ph}), a diagrammatic expansion for the quasiparticle interaction in terms 
of the nuclear potential can be obtained \cite{brown71} by performing functional derivatives with
respect to the occupation probabilities. Up to second order in perturbation theory
one obtains for a general two-body interaction $V_{2N}$:
\begin{equation}
{\mathcal F}^{(1)}_{2N}({\vec p}_1 s_1 t_1; {\vec p}_2 s_2 t_2) 
= \langle 12 | \bar V_{2N} | 12 \rangle
\label{qpipt1}
\end{equation}
\begin{equation}
{\mathcal F}^{(2pp)}_{2N} ({\vec p}_1 s_1 t_1; {\vec p}_2 s_2 t_2) 
= \frac{1}{2} \sum_{mn} \frac{|\langle 12 | \bar V_{2N} | mn \rangle|^2 \bar n_m
\bar n_n} {e_1 + e_2 - e_m - e_n}
\label{qpi2pp}
\end{equation}
\begin{equation}
{\mathcal F}^{(2hh)}_{2N} ({\vec p}_1 s_1 t_1; {\vec p}_2 s_2 t_2)
=\frac{1}{2} \sum_{ij} \frac{|\langle ij | \bar V_{2N} | 12 \rangle |^2 n_i n_j}
{e_i + e_j - e_1 - e_2}
\label{qpi2hh}
\end{equation}
\begin{equation}
{\mathcal F}^{(2ph)}_{2N} ({\vec p}_1 s_1 t_1; {\vec p}_2 s_2 t_2) 
= -2 \sum_{jn} \frac{|\langle 1j | \bar V_{2N} | 2n \rangle |^2 n_j \bar n_n}
{e_1 + e_j - e_2 - e_n},
\label{qpi2ph}
\end{equation}
which correspond respectively to diagrams (a), (b), (c), and (d) in Fig.\ \ref{diag}. The first-order contribution
in Eq.\ (\ref{qpipt1}) is just the antisymmetrized two-body potential for two nucleons restricted to the Fermi
surface. It contains only the four central terms $f, f^\prime, g, g^\prime$ as well as the two relative momentum
tensor interactions $h, h^\prime$. The second-order contributions in Eqs.\ (\ref{qpi2pp})$-$(\ref{qpi2ph})
give rise generically to the center-of-mass tensor interactions $k, k^\prime$, but only the particle-hole term 
Eq.\ (\ref{qpi2ph}) can generate the cross-vector interactions $l, l^\prime$ through the interference of a 
spin-orbit interaction with any other nonspin-orbit component in the bare nucleon-nucleon potential \cite{holt13prca}.

The expressions in 
Eqs.\ (\ref{qpipt1})$-$(\ref{qpi2ph}) can be decomposed into partial wave matrix elements of the bare nucleon-nucleon
potential or the derived medium-dependent 2N interaction. In Section \ref{pwd} below, we give explicit expressions 
for the partial-wave matrix elements of the density-dependent 2N interaction  
derived from the N2LO \cite{holt10} and N3LO \cite{kaiser18,kaiser19} chiral three-body force. To date, the 
contributions from the N2LO chiral three-body force have been included \cite{holt13prca,holt12npa,holt17}
exactly in the calculation of the quasiparticle interaction in isospin-symmetric nuclear matter and pure neutron matter.
At first order in perturbation theory, the second functional derivative of Eq.\ (\ref{e1nnn}) leads to
\begin{equation}
{\mathcal F}^{(1)}_{3N}({\vec p}_1 s_1 t_1 , {\vec p}_2 s_2 t_2) = \sum_i
 n_i \langle i12 | \bar{V}_{3N} | i12 \rangle,
\label{qpi3nb}
\end{equation}
where $\bar V_{3N}$ is the fully antisymmetrized three-body force. This is equivalent to the definition of the
density-dependent NN interaction in Eq.\ (\ref{ddnn}) but restricted by the kinematics of quasiparticles lying
on the Fermi surface. Moreover, the use of the in-medium 2N interaction constructed assuming on-shell scattering
in the center-of-mass frame is not appropriate (in particular, it would give no center-of-mass dependence at
leading order in Eq.\ (\ref{qpi3nb})). Explicit and exact expressions (in the absence of a momentum-space cutoff) 
for the Landau Fermi liquid parameters in Eqs.\ (\ref{qpi1})$-$(\ref{gflp})
from the N2LO chiral three-nucleon force have therefore been derived in Ref.\ \cite{holt17}. Only the higher-order perturbative
contributions to the quasiparticle interaction (where medium effects are included through normal Pauli blocking
of intermediate states) utilize the in-medium 2N interaction derived in the center-of-mass frame.
In the following we highlight their 
qualitative significance on the different terms of the quasiparticle interaction.

\begin{figure}[t]
\begin{center}
\includegraphics[height=20.1cm]{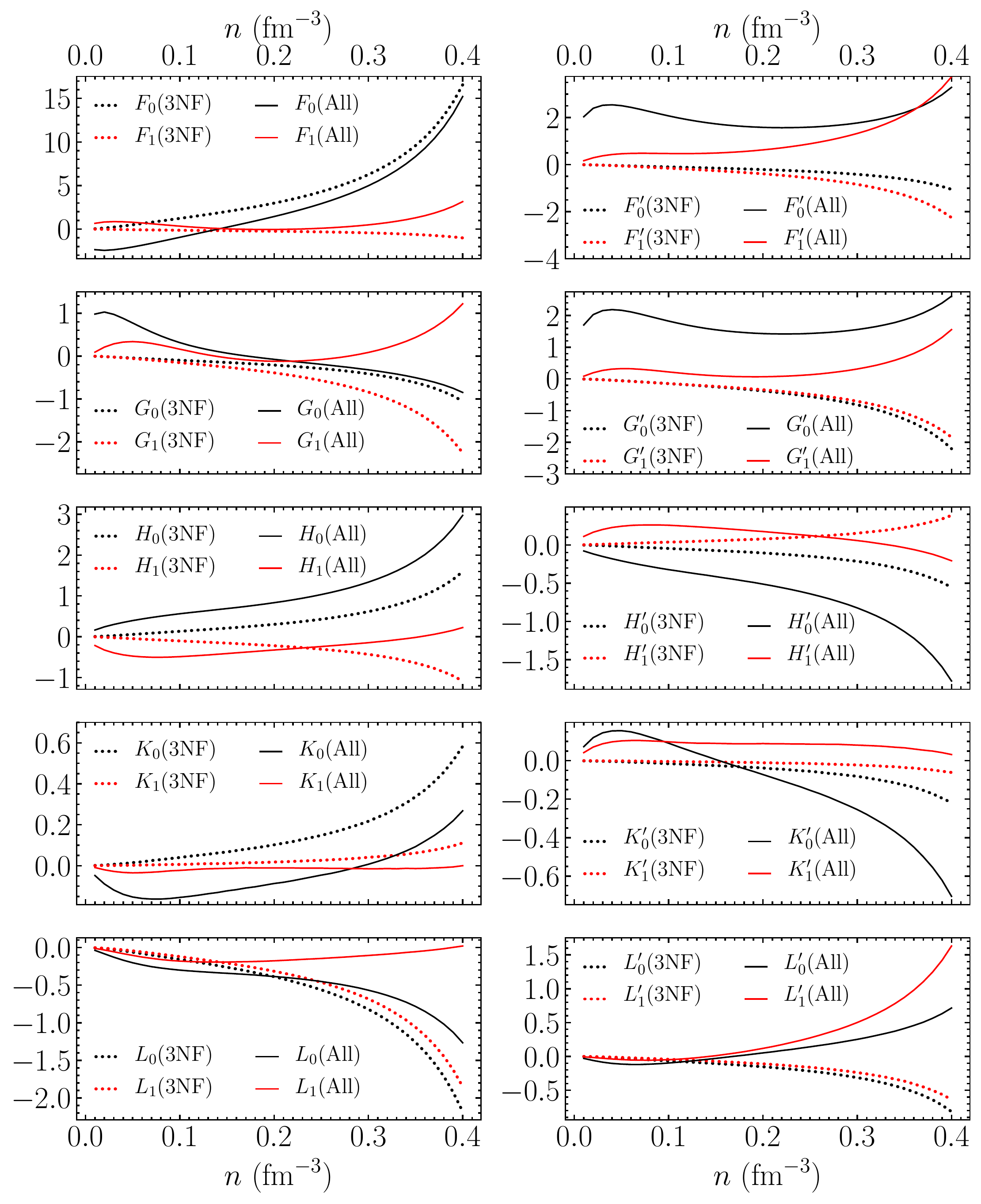}
\end{center}
\vspace{-.5cm}
\caption{Density-dependent dimensionless Fermi liquid parameters in isospin-symmetric nuclear matter. Dotted lines
symbolize the first-order perturbative contribution from three-body forces, while solid lines represent the sum of 
all second-order contributions including two- and three-body forces.}
\label{FLP}
\end{figure}

In Fig.\ \ref{FLP} we plot the dimensionless Fermi liquid parameters associated with the $L=0,1$ 
Legendre polynomials (black and red dotted lines respectively) in isospin-symmetric nuclear 
matter from the N2LO chiral three-body force as a function of the nucleon density (up to $n=0.4$\,fm$^{-3}$). 
Although one may be skeptical of results from chiral effective field theory beyond $n \simeq 2n_0$, the
Landau parameters must obey stability inequalities, e.g.,
\begin{equation}
Q_L > -(2L+1),
\label{stab}
\end{equation}
where $Q \in \{F,F^\prime,G,G^\prime\}$, for the central components of the quasiparticle
interaction. Therefore we find it informative to speculate on the high-density behavior of the Landau 
parameters, since they might give hints toward possible instability mechanisms in dense matter.
We note that complete stability conditions involving all spin-dependent interactions $H, K, L$ 
(and $H^\prime, K^\prime, L^\prime$) that couple to $G$ (and $G^\prime$) have not yet been worked out.
To date only the effect of the relative tensor quasiparticle interaction has been 
considered \cite{backman79}. We have found that in the presence of such
Pomeranchuk instabilities, perturbation theory itself can be poorly behaved. For instance, in symmetric nuclear
matter with density $n \lesssim n_0/2$ (where $F_0 < -1$ and nuclear matter is unstable to density fluctuations), 
we have computed also the third-order particle-particle 
contributions to the Fermi liquid parameters and found that $F_0$ is of comparable 
size to the second-order particle-particle diagrams. For other Fermi liquid parameters, however, the third-order
particle-particle contributions are generally small at low densities.

The dotted lines in Fig.\ \ref{FLP} are obtained from only the leading contribution due to three-body forces in Eq.\ (\ref{qpi3nb}).
The solid lines represent the Fermi liquid parameters obtained from the sum of two- and three-body forces up to
second order in perturbation theory. For the second-order contributions in Eqs.\ (\ref{qpi2pp})$-$(\ref{qpi2ph}) we
have replaced the 
two-body interaction $V_{2N}$ with $V_{2N} + V_{\rm med}$, where $V_{2N}$ is the N3LO-450 
potential and $V_{\rm med}$ is the consistent density-dependent interaction constructed from the N2LO 
three-body force with nonlocal regulator. For several Fermi liquid parameters, we see that three-body forces
provide the dominant contribution at high density.
For instance, the strong increase in the $F_0$ Landau parameter (top left panel
of Fig.\ \ref{FLP}) as a function of density is a direct result of the first-order contribution from three-body forces. 
The nuclear matter incompressibility ${\mathcal K} = 9 \partial P / \partial \rho$,
where $P = \rho^2 \frac{\partial (E/A)}{\partial \rho}$, is related to the $F_0$ Landau parameter through
\begin{equation}
{\mathcal K}=\frac{3k_f^2}{M^*} \left (1+F_0\right ),
\label{comp}
\end{equation}
where $M^*$ is the nucleon effective mass, and 
therefore three-body forces play a central role in the saturation mechanism \cite{bogner05} of nuclear matter
with chiral nuclear forces.
On the other hand, in some cases three-body
forces play only a minor role, such as for the Landau parameters $F_1$ and $F_0^\prime$. The former
is related to the nucleon effective mass through
\begin{equation}
\frac{M^*}{M} = 1+\frac{F_1}{3},
\label{effmass}
\end{equation}
and the latter is related to the nuclear isospin-asymmetry energy through
\begin{equation}
S_2=\frac{k_f^2}{6M^*} \left (1+F_0^\prime \right ),
\label{comp}
\end{equation}
where $S_2$ is defined as the first term in a power series expansion of the nuclear equation of state about 
the isospin-symmetric configuration:
\begin{equation}
\frac{E}{A}(n,\delta_{np}) = \frac{E}{A}(n,0) + S_2(n)\delta_{np}^2 + \cdots
\label{esyme}
\end{equation}
with $\delta_{np} = \frac{n_n-n_p}{n_n+n_p}$.

In general, we see from Fig.\ \ref{FLP} that the noncentral components $K$ and $L$ 
of the quasiparticle interaction that depend explicitly on the center-of-mass momentum $\vec P$
are small at nuclear saturation density. However, several of the associated Fermi liquid parameters,
such as $K_0^\prime$, $L_0$, and $L_1^\prime$ begin to grow rapidly for higher densities. Therefore,
even though there has been little motivation to include such terms in modern energy density functionals
fitted to the properties of finite nuclei, such novel interactions may become more relevant in applications
related to neutron star physics. The full quasiparticle interaction in pure neutron matter has already
been computed \cite{holt13prca} with modern chiral two- and three-nucleon forces. One finds again an 
enhanced role of three-body forces on the incompressibility of pure neutron matter 
and therefore the stability of neutron stars against gravitational collapse. The more general
case of the quasiparticle interaction in nuclear matter at arbitrary isospin-asymmetry is in progress.

\begin{figure}[t]
\begin{center}
\includegraphics[height=3.5cm]{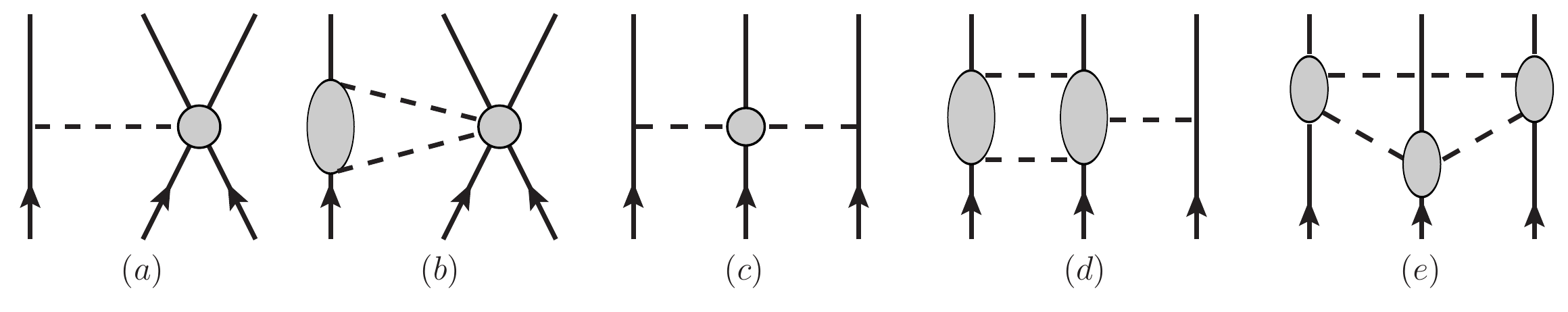}
\end{center}
\caption{Schematic representation of the diagrammatic contributions to the chiral three-nucleon force at 
next-to-next-to-next-to-leading order (N3LO).}
\label{3NF_N3LO}
\end{figure}

\section{Chiral three-nucleon force at next-to-next-to-next-to-leading order}

Up to now we have considered only the chiral three-body force at N2LO.
At order N3LO in the chiral power counting,
additional three- and four-nucleon forces arise without any additional undetermined low-energy
constants. However, except in the case of pure neutron matter, the inclusion of the N3LO three-body contributions
requires a refitting of the three-body low-energy constants $c_D$ and $c_E$. 
The N3LO three-body force is written schematically as
\begin{equation}
V_{3N}^{(4)} = V_{1\pi-\rm cont.}^{(4)} + V_{2\pi-\rm cont.}^{(4)} + V_{1/M}^{(4)} + V_{2\pi}^{(4)} 
+ V_{2\pi-1\pi}^{(4)}+V_{\rm ring}^{(4)},
\end{equation}
corresponding to the $1\pi-{\rm contact}$, $2\pi-{\rm contact}$, relativistic $1/M$, $2\pi$, $2\pi-1\pi$, and ring topologies, 
respectively. All contributions have been
worked out and presented in Refs.\ \cite{ishikawa07,bernard08,bernard11}. Although we will not consider their 
specific effects in the present work, we note that the leading four-nucleon forces 
have been calculated in Ref.\ \cite{epelbaum06}. In deriving the density-dependent 2N interaction at 
N3LO, we take the expressions
from Refs.\ \cite{bernard08,bernard11} based on the method of unitary transformations.

The density-dependent 2N interaction from the short-range terms and relativistic corrections, shown diagrammatically 
in Fig.\ \ref{3NF_N3LO}(a) and (b), was computed first 
in Ref.\ \cite{kaiser18}. Results were derived in the absence of a regulating function depending explicitly on the
value of the intermediate-state momentum $k_3$ in Eq.\ (\ref{ddnn}). The resulting expressions for $V_{\rm med}$
obtained from the N3LO 3N force could therefore be simplified to analytical expressions involving at most a 
one-dimensional integration. In Ref.\ \cite{kaiser18} it was found that the $1\pi$-exchange contact topology
proportional to the 2N low-energy constant $C_T$ gives rise to a vanishing contribution to $V_{\rm med}$ in 
isospin-symmetric nuclear matter. The density-dependent 2N interaction derived from the long-range contributions 
to the N3LO three-body force, shown diagrammatically in Fig.\ \ref{3NF_N3LO}(c), (d), and (e), was calculated in 
Ref.\ \cite{kaiser19}. Again, the integration over the three-dimensional filled Fermi sphere could be performed up 
to at most one remaining integration. The formulas for the density-dependent NN interaction from the N3LO
three-body force are quite lengthy, and we refer the reader to Refs.\ \cite{kaiser18,kaiser19} for additional details.

\subsection{Partial-wave decomposition}
\label{pwd}
The analytical expressions for the medium-dependent 2N potential $V_{\rm med}$ obtained from the
N3LO chiral three-body force \cite{kaiser18,kaiser19} 
can be conveniently understood by examining their attractive or repulsive effects 
in various partial waves. For comparison we will show also the lowest-order partial-wave contributions 
from the N2LO chiral three-body force, however, we note that the values of the three-body contact
terms will need to be refitted in order to make a consistent comparison. In all cases, we choose the values 
$c_1 = -0.81$\,GeV$^{-1}$, $c_3=-3.4$\,GeV$^{-1}$, 
$c_4=3.4$\,GeV$^{-1}$, $c_D=-0.24$, and $c_E=-0.106$, which have been used in other calculations presented
in this work. We recall that the low-energy constants $c_D$ and $c_E$ 
of the N2LO chiral 3N force are fitted (including the N3LO 
chiral 2N interaction with cutoff scale $\Lambda = 450$\,MeV) to
the binding energies of $^3$H and $^3$He as well as the beta-decay lifetime of $^3$H. Comparing to the values
of $c_D$ and $c_E$ fitted in combination with the N2LO two-body force (see Table II of Ref.\ \cite{sammarruca15}), 
we do not expect qualitative differences in the results below coming from these two different choices in the chiral order.
For the leading-order (LO) contact term $C_T$ that
appears in the $1\pi$- and $2\pi$-contact topologies, we use the value $C_T = -2.46491$\,GeV$^{-2}$ from
the N3LO-450 2N potential.

We follow the description in Ref.\ \cite{erkelenz71} to obtain the diagonal momentum-space 
partial-wave matrix elements of the density-dependent NN interaction. With start with the form of a general 
nucleon-nucleon potential:
\begin{eqnarray}
&&\hspace{-.5in}V(\vec p,\vec q\,) = V_C + \vec \tau_1 \cdot \vec \tau_2\, W_C + 
\left [V_S + \vec \tau_1 \cdot \vec \tau_2 \, W_S \right ] \vec \sigma_1 \cdot 
\vec \sigma_2 + \left [ V_T + \vec \tau_1 \cdot \vec \tau_2 \, W_T \right ] 
\vec \sigma_1 \cdot \vec q \, \vec \sigma_2 \cdot \vec q \nonumber \\
&&\hspace{-.2in}+ \left [ V_{SO} + \vec \tau_1 \cdot \vec \tau_2 \, W_{SO} \right ] \,
i (\vec \sigma_1 + \vec \sigma_2 ) \cdot (\vec q \times \vec p\,)
+ \left [ V_{Q} + \vec \tau_1 \cdot \vec \tau_2 \, W_{Q} \right ] \,
\vec \sigma_1 \cdot (\vec q \times \vec p\,)\, \vec \sigma_2 \cdot (\vec q \times
\vec p\,),
\end{eqnarray}
where the subscripts refer to the central ($C$), spin-spin ($S$), tensor ($T$), 
spin-orbit ($SO$), and quadratic spin orbit ($Q$) components, each with an 
isoscalar ($V$) and isovector ($W$) version. The diagonal (in momentum space) 
partial-wave matrix elements for 
different spin and orbital angular momentum channels are then 
given in terms of the functions
$U_K = V_K+(4I-3) W_K$, where $K \in \{C,S,T,SO,Q\}$ and
the total isospin quantum number takes the values $I=0,1$. Explicit expressions
can be found in Ref.\ \cite{holt10}.

\begin{figure}[t]
\begin{center}
\includegraphics[height=9.4cm]{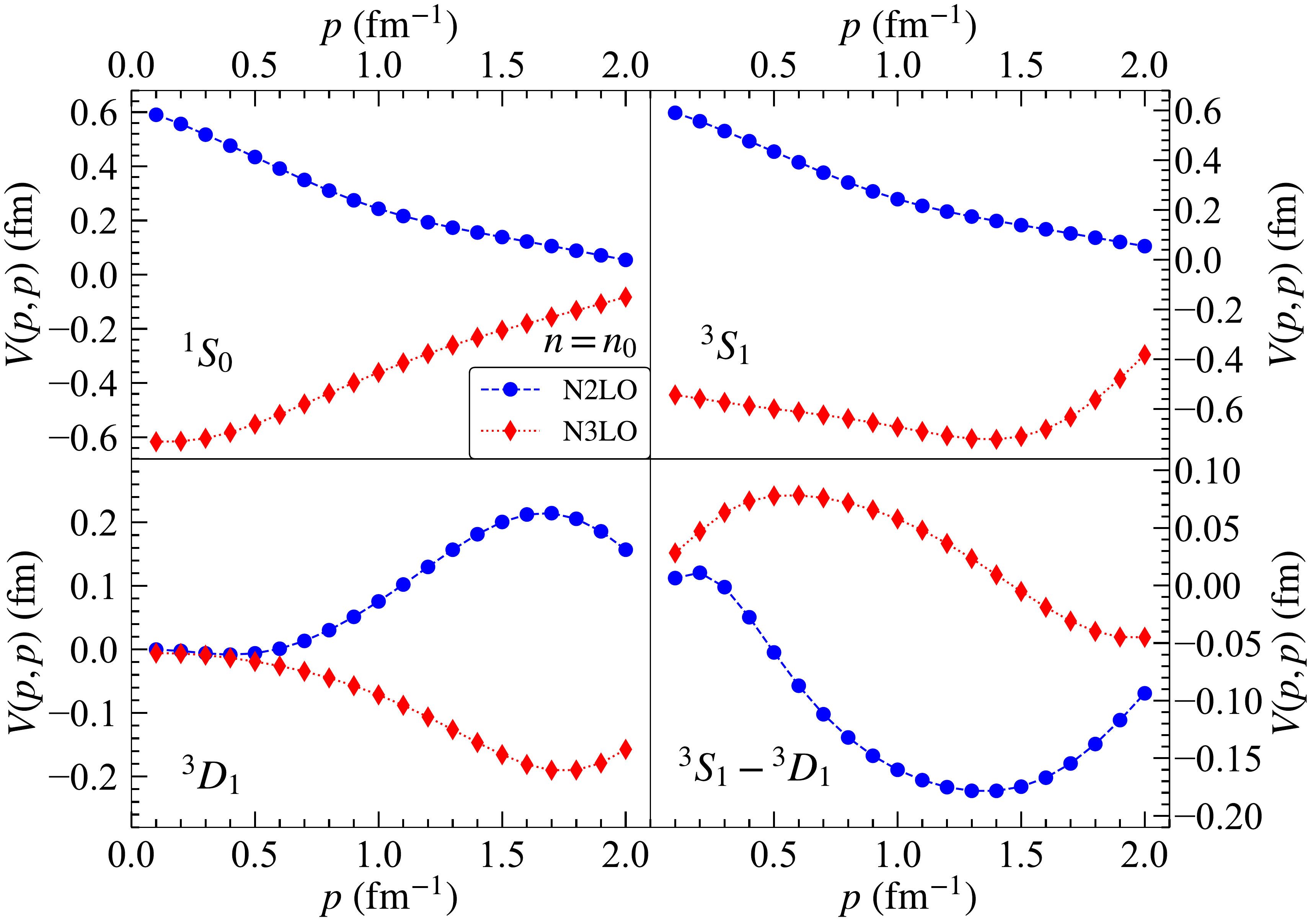}
\end{center}
\caption{Diagonal momentum-space matrix elements of $V_{\rm med}$ associated with the 
total N2LO and N3LO three-body force in the 
$^1S_0$ and $^3S_1-{^3D_1}$ partial-wave channels at $n = n_0$ in isospin-symmetric nuclear 
matter.}
\label{fig:SD133}
\end{figure}

\begin{figure}[t]
\begin{center}
\includegraphics[height=9.4cm]{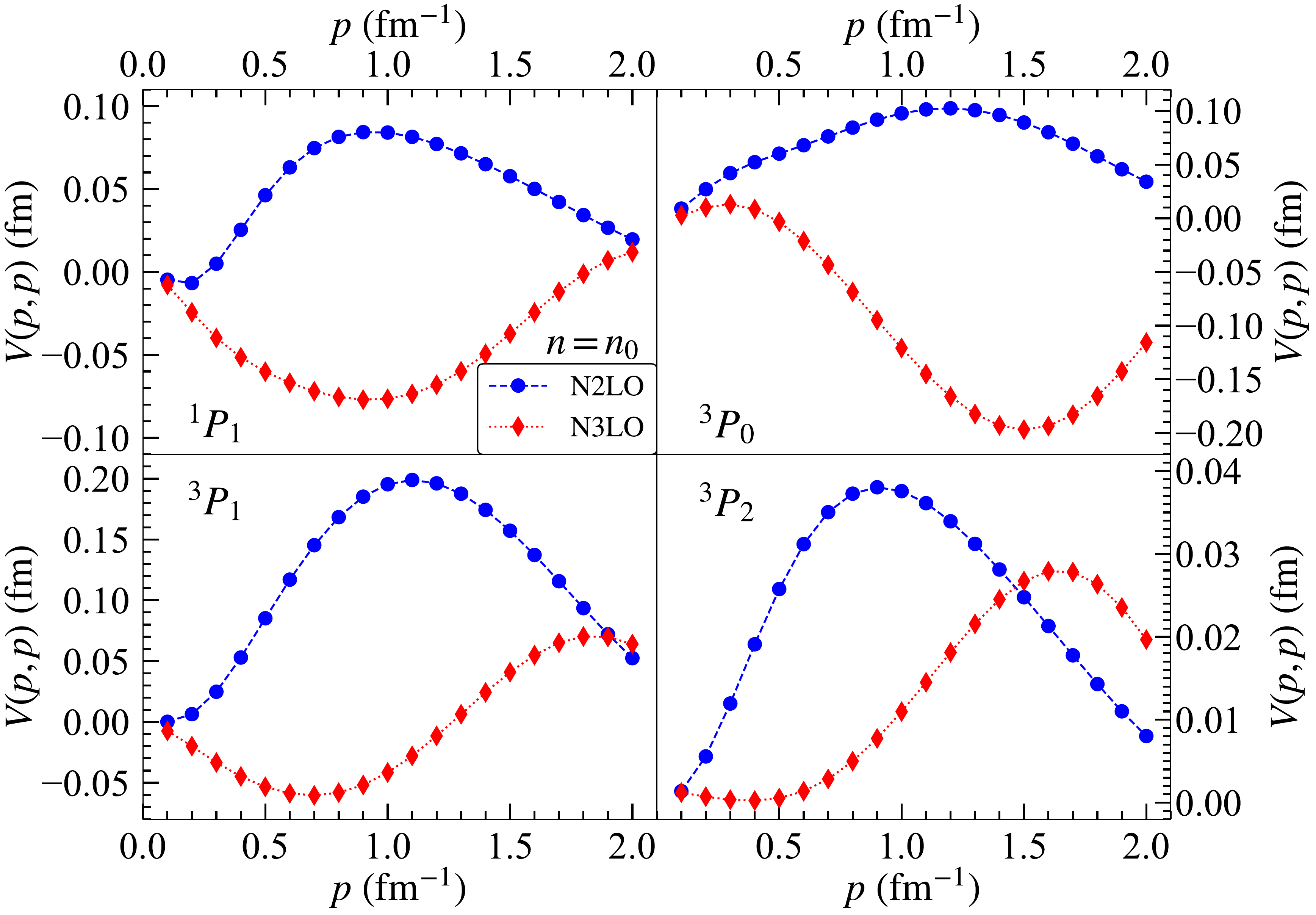}
\end{center}
\caption{Diagonal momentum-space matrix elements of $V_{\rm med}$ associated with the total N2LO 
and N3LO three-body force in the $^1P_1$, $^3P_0$, $^3P_1$, and $^3P_2$ partial-wave channels 
at $n = n_0$ in isospin-symmetric nuclear matter.}
\label{fig:P133}
\end{figure}

In Fig.\ \ref{fig:SD133} we show the $^1S_0$, $^3S_1$, $^3D_1$, $^3S_1 - {^3D_1}$ diagonal momentum-space matrix
elements of $V_{\rm med}$ from the N2LO (blue circles) and N3LO (red diamonds) chiral three-nucleon 
force in isospin-symmetric nuclear matter at the density $n = n_0$. Note that we have multiplied the matrix
elements by the nucleon mass $M$ to obtain dimensions of [fm]. Interestingly, we observe that 
the total N3LO three-body force in these partial-wave channels is roughly equal in magnitude but 
opposite in sign compared to the N2LO three-body force. Whereas the N2LO three-body force is
largely repulsive in symmetric nuclear matter at saturation density, the N3LO three-body force is
strongly attractive, except in the case of the coupled $^3S_1-{^3D_1}$ tensor channel. 
One should keep in mind, however, that the low-energy constants $c_D$ and $c_E$
must be refitted after the introduction of the N3LO three-body force. One might expect from the above
observations that the N2LO three-body force would be enhanced in order to offset the opposite behavior
introduced from the N3LO three-body force in the lowest partial-wave channels. 

In Fig.\ \ref{fig:P133} we show the $^1P_1$, $^3P_0$, $^3P_1$, and $^3P_2$ diagonal momentum-space matrix
elements of $V_{\rm med}$ from the N2LO (blue circles) and N3LO (red diamonds) chiral three-nucleon 
force in isospin-symmetric nuclear matter at the density $n = n_0$. In both the $^1P_1$ and $^3P_0$
channels, the N2LO and N3LO contributions are approximately equal in magnitude but opposite in sign.
In the $^3P_1$ channel, which is repulsive in the bare 2N potential, we see that the combination of N2LO and
N3LO contributions enhances the repulsion. The $^3P_2$ channel, which is attractive in the free-space 2N potential,
also receives repulsive contributions from the N2LO and N3LO in-medium interaction $V_{\rm med}$.
The feature that N3LO loop corrections are not small compared to N2LO tree contributions has been 
seen in several instances, e.g., in pion-nucleon scattering \cite{bernard97} as well as the three-nucleon force 
derivation \cite{bernard08} and its application \cite{golak14}. The results presented in 
Figs.\ \ref{fig:SD133} and \ref{fig:P133} are state of the art and may change 
if the effects of sub-sub-leading chiral 3N forces are included. For the $2\pi-1\pi$ and ring topologies, the 
N4LO corrections are sizable and dominate in most cases over the nominally leading N3LO terms \cite{krebs13}.

\begin{figure}[t]
\begin{center}
\includegraphics[height=7.5cm]{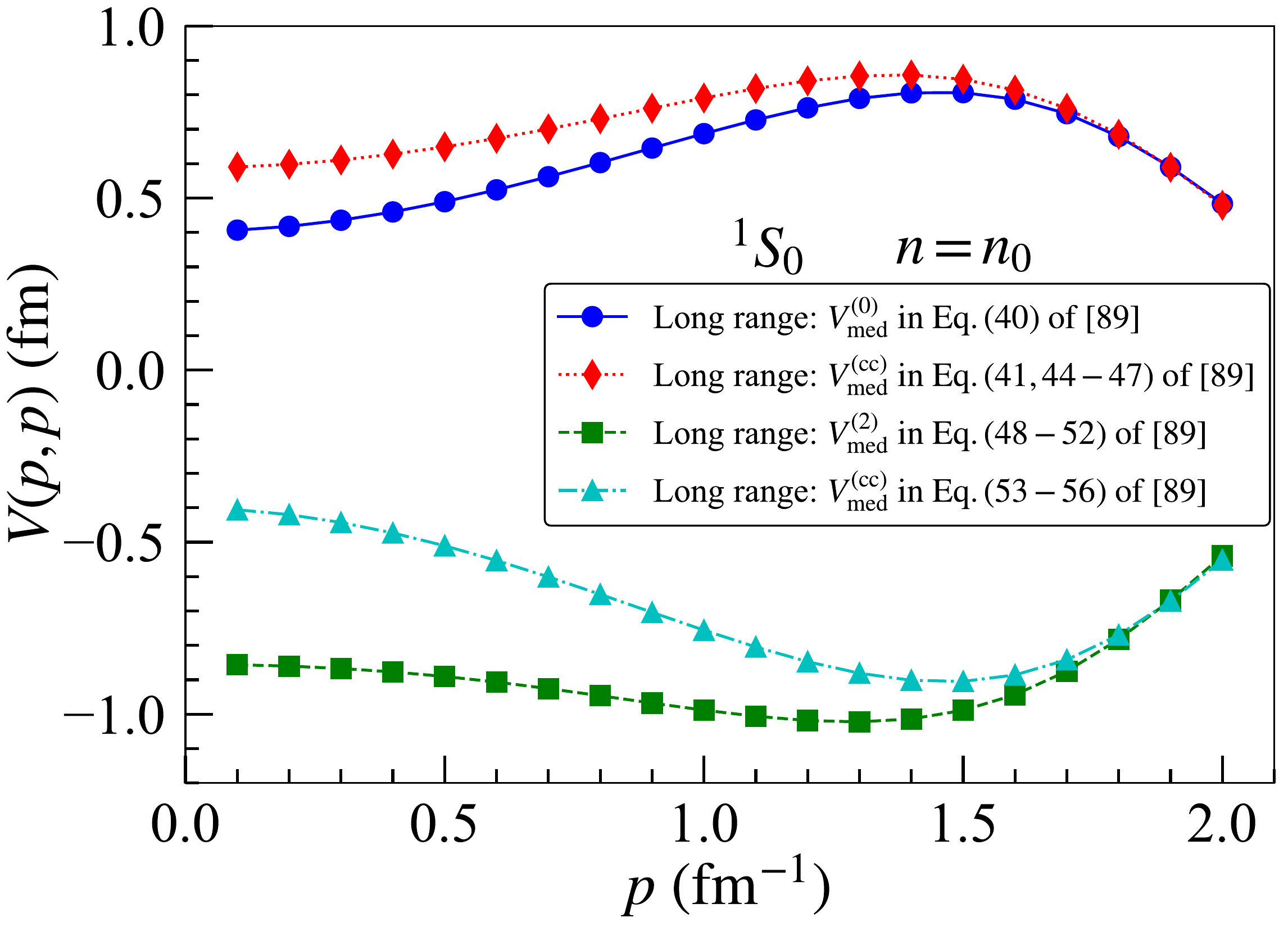}
\end{center}
\caption{Diagonal momentum-space matrix elements of the N3LO three-body force for selected topologies in the 
$^1S_0$ partial-wave channel at saturation density $n_0$ in isospin-symmetric nuclear matter.}
\label{fig:1S0}
\end{figure}

\begin{figure}[t]
\begin{center}
\includegraphics[height=7.cm]{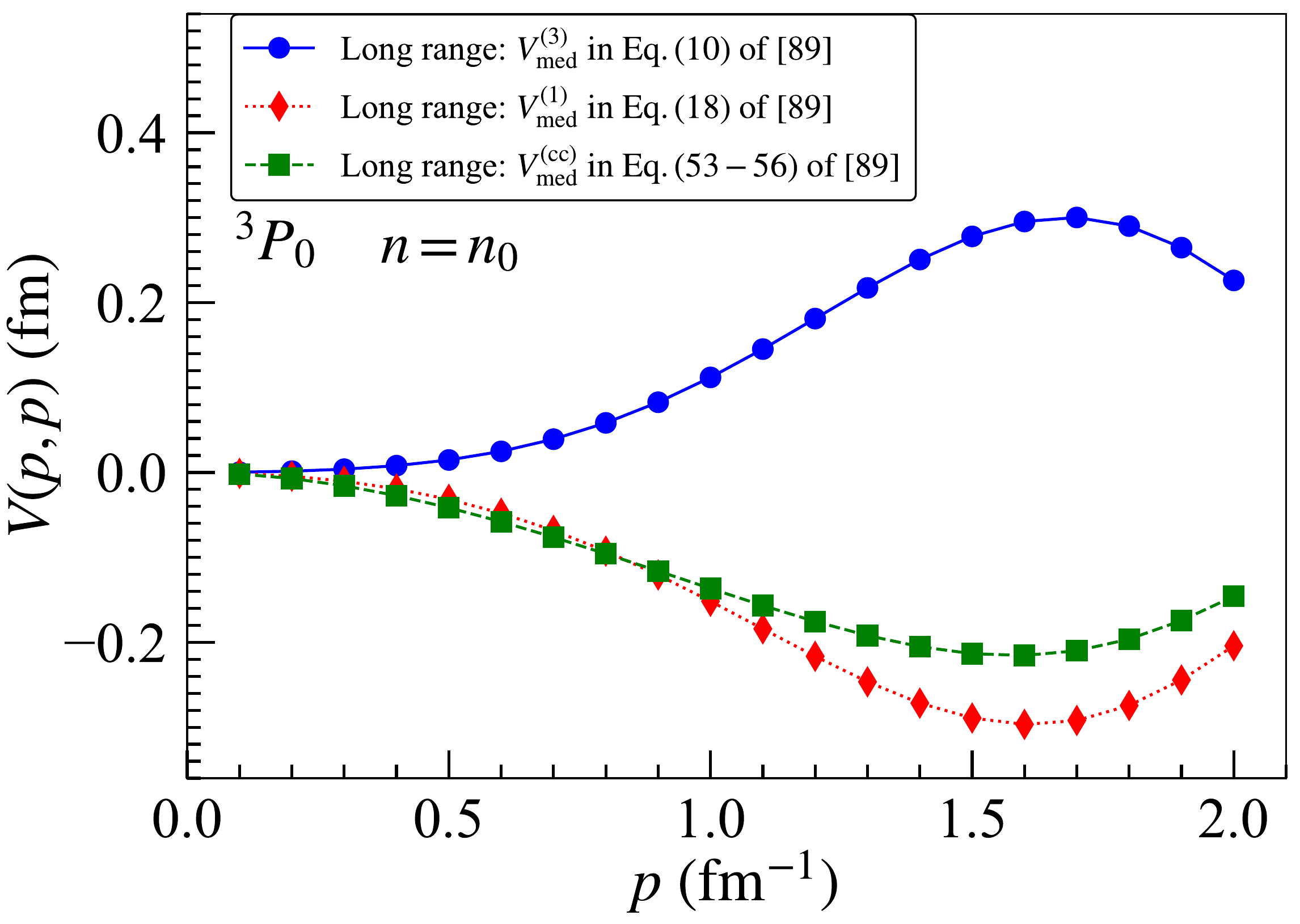}
\end{center}
\caption{Diagonal momentum-space matrix elements of the N3LO three-body force for selected topologies in the 
$^3P_0$ partial-wave channel at saturation density $n_0$ in isospin-symmetric nuclear matter.}
\label{fig:3P0}
\end{figure}

The long-range parts of the N3LO chiral three-body force are expected \cite{tews13} to give larger contributions
to the equation of state than the relativistic $1/M$ corrections and the $2\pi-$contact topologies. Due to the 
large number of contributions to $V_{\rm med}$ at N3LO, we only show selected results for individual topologies.
In Fig.\ \ref{fig:1S0} we plot several of the dominant pion-ring contributions to the $^1S_0$ partial-wave matrix elements
for the density-dependent NN interaction derived from the N3LO chiral three-body force. We see that individual 
long-range contributions are large, but sizable cancellations lead to an overall reduced attractive
$^1S_0$ partial-wave channel at low momenta. In Fig.\ \ref{fig:3P0} we plot several of the important
$2\pi$, $2\pi$-$1\pi$, and ring topology contributions
to the $^3P_0$ partial-wave matrix elements of the density-dependent NN interaction derived from the N3LO chiral 
three-body force. We again find large cancellations among individual terms, but the sum produces significant
attraction in this partial-wave channel.

\section{Summary and Conclusions}

We have reviewed the construction and implementation of density-dependent two-body interactions
from three-body forces at N2LO and N3LO in the chiral expansion. We showed that at leading order
in many-body perturbation theory, the in-medium 2N interaction reproduces very well the exact 
contributions to the nuclear equation of state and nucleon self energy from the complete three-body
force. The standard nonlocal high-momentum regulator used in our previous works leads to simpler
analytical expressions for the density-dependent 2N interaction, consistency with the bare 2N potential, 
and relatively small artifacts in both the equation of state up to twice saturation density and the 
single-particle potential up to $p \simeq 400-500$\,MeV at nuclear matter saturation density. Local 3N 
regulators with the same value of the cutoff, $\Lambda_{\rm loc} = \Lambda_{\rm nonloc}$, have been 
commonly used in previous studies of nuclear few-body systems, but these are shown to produce very 
large artifacts, even in the nuclear equation of state at saturation density. This could be remedied by 
choosing a local regulating function with $\Lambda_{\rm loc} = 2 \Lambda_{\rm nonloc}$, which is well 
motivated since the momentum transfer $q$ can reach values twice as large as the relative momentum
for two particles on the Fermi surface.

The use of medium-dependent two-body interactions has been shown to facilitate the implementation
of three-body forces in higher-order perturbative calculations of the nuclear equation of state, single-particle
potential, and quasiparticle interaction. In particular, nuclear matter was shown to saturate at the correct binding
energy and density within theoretical uncertainties when computed up to third-order in perturbation theory. 
Moreover, microscopic nucleon-nucleus optical potentials 
derived from chiral two- and three-body forces have been shown to accurately predict proton elastic
scattering cross sections on calcium isotopes up to projectile energies of $E\simeq 150$\,MeV. 
The use of medium-dependent NN potentials derived from the N3LO chiral three-body force for calculations
of the nuclear equation of state, single-particle potential, and quasiparticle interaction remain a topic of future
research. As a first step, we have performed a partial-wave decomposition of $V_{\rm med}$ at N3LO in the 
chiral expansion and shown that the effective interaction is expected to be attractive in symmetric nuclear
matter around saturation density.


\section*{Conflict of Interest Statement}

The authors declare that the research was conducted in the absence of any commercial or financial relationships that could be construed as a potential conflict of interest.


\section*{Funding}
Work of J.\ W.\ H.\ is supported by the National Science Foundation under Grant No.\ PHY1652199 and by the U.S.\ 
Department of Energy National Nuclear Security Administration under Grant No.\ DE-NA0003841. The work of M.\ K.\ 
was supported in part by JSPS Grant-in-Aid for JSPS Research Fellow No.\ 18J15329. The work of N.\ K.\ has been 
supported in part by DFG and NSFC (CRC110).

\section*{Acknowledgments}
Portions of this research were conducted with the advanced computing resources provided by Texas A\&M High Performance Research Computing. This manuscript has been released as a Pre-Print at \cite{holt19}.



\bibliographystyle{frontiersinHLTH&FPHY} 


\end{document}